# Non-modal stability analysis of stratified two-phase channel flows

I. Barmak[a)], A. Yu. Gelfgat, A. Ullmann, and N. Brauner

*School of Mechanical Engineering, Tel Aviv University, Tel Aviv 69978, Israel*

The non-modal transient growth of perturbations in horizontal and inclined channel flows of two immiscible fluids is studied. 3D perturbations are examined in order to find the optimal perturbations that attain the maximum amplification of perturbation energy at relatively short times. Definition of the energy norm is extended to account for the gravitational potential energy along with the kinetic energy and interfacial capillary energy. Contrarily to the fastest exponential growth, which is reached by essentially 2D perturbations, the maximal non-modal energy growth is attained mostly by three-dimensional spanwise perturbations. Significant transient energy growth is found to occur in linearly stable flow configurations, which, similarly to single phase shear flows, may trigger non-linear destabilizing mechanisms within one of the phases. It is shown that the transient energy growth in linearly stable cases can be accompanied by noticeable interface deformations. Therefore, flow pattern transition due to non-modal transient growth and reduction of the range of operational conditions for which stratified-smooth flow remains stable cannot be ruled out.

*Keywords*: Two-phase stratified flow; non-modal stability; transient energy growth; optimal perturbations; gravity effect.

## 1. INTRODUCTION

Linear stability of stratified two-phase flows in horizontal and inclined channels was studied by Yih(1967), Hooper and Boyd (1983), Yiantsios and Higgins (1988), Charru and Fabre (1994), Tilley et al. (1994), Ó Náraigh et al. (2014), Kaffel and Riaz (2015), and others. Recently, this problem was addressed more rigorously with reference to the prediction of the operational region corresponding to stable stratified-smooth flow on flow pattern maps (Barmak et al., 2016a, b). Those studies used the traditional (modal) approach that is based on the investigation of the eigenvalue problem, in which eigenvectors and eigenvalues represent perturbation amplitudes and their corresponding growth rate. Upon specifying the wavenumber, the same exponential growth of perturbations of all flow variables is assumed. The maximal perturbation amplitude can be initiated either in the bulk of one of the phases or at the interface. In any case, it is the growth of the interface displacement amplitude that is responsible for flow pattern transition from stratified-smooth flow to other flow patterns (e.g., stratified-wavy, plug/slug flow). On a flow stability map, a neutral stability boundary of the modal analysis defines the critical conditions and the associated critical wave number, as well as the linearly stable (subcritical) region of operational conditions for which all perturbations eventually decay exponentially with time. In that region, stratified flow with a smooth interface is expected to be stable (Barmak et al., 2016a,b).

The present study examines the growth of initially small 2D and 3D perturbations at relatively short times at which the energy of perturbations can grow substantially even in the subcritical regime, before starting to decay exponentially in time. This phenomenon is known as non-modal instability, and was studied in detail for single-phase Couette and Poiseuille shear flows (Reddy and Henningson, 1993; Schmid and Henningson,

---

[a)] Author to whom correspondence should be addressed. Electronic mail: ilyab@tauex.tau.ac.il.

2001). The transient energy growth was argued to be responsible for transition from laminar flow at subcritical values of the Reynolds number (Brandt 2014). In the unstable regime, it was shown that the non-modal effects can enhance perturbation energy significantly before modal behavior dominates, and therefore can be considered important for studying transition scenarios (e.g., Lucas et al., 2015, Jose et al., 2017).

Similar transient energy growth can be observed also in two-phase flows, however only a few studies addressed the issue. Van Noorden et al. (1998) and South and Hooper (1999) examined the non-modal growth of two-dimensional perturbations. Yecko (2008) considered three-dimensional perturbations and investigated the role of capillarity (due to interfacial tension) on the transient energy growth in the case of sheared fluids of similar densities, while gravity effects were neglected and the Froude number was not introduced. Following the discussion of Renardy (1987) and South and Hooper (1999) about a valid energy norm for stratified two-phase flow, the one composed of the kinetic energy and the interfacial energy was chosen as a perturbation measure. The transient energy growth at short times is attributed to formation of streamwise streaks evolving from initial streamwise growth of vortices via the lift-up effect. Such an algebraic instability was originally observed in single-phase flows (Schmid and Henningson, 2001) and later found also in two-phase mixing layers (Yecko and Zaleski, 2005; Malik and Hooper, 2007).

In the present study, we extend the non-modal analysis to examine the growth of 2D and 3D perturbations in horizontal and inclined flows of two fluids of different densities and viscosities in the gravitational field. Accordingly, the gravitational potential energy is added to the definition of the energy norm. We are looking for the so-called optimal perturbations (Farrell, 1988) that exhibit the maximum gain for energy transfer from the mean flow to the perturbations. These perturbations may trigger nonlinear effects that cause instability of stratified two-phase flows also under subcritical conditions obtained by the modal analysis. The non-modal growth of the perturbation energy can be associated with perturbation growth in the bulk of one or both of the phases, and/or with growth of the interface displacement amplitude. While the former is addressed in the literature, the latter has never been examined. These issues, which are evidently important for the prediction of the stratified–smooth flow boundaries and flow pattern transitions in two-phase flows, are elaborated in detail in the current study.

The problem formulation for 3D perturbations in stratified flow of two immiscible fluids is presented in section 2, and is followed by the non-modal stability analysis (section 3). The analysis is applied first to zero-gravity systems (section 4A) and validated for the case study of Yecko (2008), which is shown, according to the modal analysis, to correspond to supercritical (unstable) conditions. Along with the successful comparison, we consider several examples of non-modal perturbation growth in the subcritical regime, where the non-modality can be the main driving mechanism for destabilization of the flow. Then we study the effect of gravity on the non-modal perturbation growth in stratified two-phase flows (section 4B). The evolution of the optimal perturbations in time and space is discussed, being interested in particular in cases where the fluid-fluid interface exhibits large deformations. Finally, we address two-phase flows in inclined channels (section 4C). These flows are predicted to be stable by the modal analysis in a relatively small range of the governing parameters. At the same time the non-modal growth there can be substantial, which is a clear effect of the additional gravitational forcing. In inclined flows, we address also a possibility of multiple base flow states for the same operational conditions (e.g., Ullmann et al., 2003a, b) and study the non-modal growth of each of them.

## 2. PROBLEM FORMULATION

The flow configuration of a stratified two-layer flow of two immiscible incompressible fluids in an inclined channel $(0 \leq \beta < \pi/2)$ is sketched in Figure 1. The flow, assumed isothermal, is driven by an imposed pressure gradient and gravity. The interface between fluids, labeled as $j = 1, 2$ (1 – lower phase, 2 – upper phase), is assumed to be flat in the undisturbed base flow state. Under this assumption, the position of the interface is obtained as additional unknown value of the steady state plane-parallel solution (see below). The flow in each liquid is described by the continuity and momentum equations that are rendered dimensionless in the standard manner (see Kushnir et al., 2014), choosing for the scales of length and velocity the height of the upper layer $h_2$ and the interfacial velocity $u_i$, respectively. The time and the pressure are scaled by $h_2/u_i$, and $\rho_2 u_i^2$, respectively.

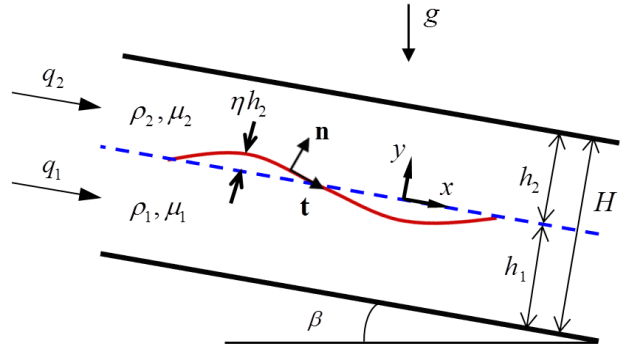

FIG. 1. Configuration of a stratified two-layer channel flow (*z*-axis comes out of the page).

For the indicated three-dimensional coordinate system (where z comes out of the page), the dimensionless continuity and momentum equations governing the flow are:

$$\frac{\partial u_j}{\partial x} + \frac{\partial v_j}{\partial y} + \frac{\partial w_j}{\partial z} = 0,$$

$$\frac{\partial u_j}{\partial t} + u_j \frac{\partial u_j}{\partial x} + v_j \frac{\partial u_j}{\partial y} + w_j \frac{\partial u_j}{\partial z} = -\frac{\rho_2}{\rho_j}\frac{\partial p_j}{\partial x} + \frac{1}{\text{Re}_2}\frac{\rho_2}{\rho_j}\frac{\mu_j}{\mu_2}\left(\frac{\partial^2 u_j}{\partial x^2} + \frac{\partial^2 u_j}{\partial y^2} + \frac{\partial^2 u_j}{\partial z^2}\right) + \frac{\sin\beta}{\text{Fr}_2},$$

$$\frac{\partial v_j}{\partial t} + u_j \frac{\partial v_j}{\partial x} + v_j \frac{\partial v_j}{\partial y} + w_j \frac{\partial v_j}{\partial z} = -\frac{\rho_2}{\rho_j}\frac{\partial p_j}{\partial y} + \frac{1}{\text{Re}_2}\frac{\rho_2}{\rho_j}\frac{\mu_j}{\mu_2}\left(\frac{\partial^2 v_j}{\partial x^2} + \frac{\partial^2 v_j}{\partial y^2} + \frac{\partial^2 v_j}{\partial z^2}\right) - \frac{\cos\beta}{\text{Fr}_2},$$

$$\frac{\partial w_j}{\partial t} + u_j \frac{\partial w_j}{\partial x} + v_j \frac{\partial w_j}{\partial y} + w_j \frac{\partial w_j}{\partial z} = -\frac{\rho_2}{\rho_j}\frac{\partial p_j}{\partial z} + \frac{1}{\text{Re}_2}\frac{\rho_2}{\rho_j}\frac{\mu_j}{\mu_2}\left(\frac{\partial^2 w_j}{\partial x^2} + \frac{\partial^2 w_j}{\partial y^2} + \frac{\partial^2 w_j}{\partial z^2}\right),$$

(1)

where $\mathbf{u}_j = (u_j, v_j, w_j)$ and $p_j$ are the velocity and pressure of the fluid $j$, $\rho_j$ and $\mu_j$ are the corresponding density and dynamic viscosity. In the dimensionless formulation the lower and upper phases occupy the regions $-n \leq y \leq 0$, and $0 \leq y \leq 1$, respectively, $n = h_1/h_2$ ($h = h_1/H = n/(n+1)$ is the lower phase holdup). Other dimensionless parameters: $\text{Re}_2 = \rho_2 u_i h_2/\mu_2$ ($\text{Re}_1 = \text{Re}_2 \, r/m$) is the Reynolds number defined using the properties of fluid 2, $\text{Fr}_2 = u_i^2/gh_2$ is the Froude number, $r = \rho_1/\rho_2$ and $m = \mu_1/\mu_2$ are the density and viscosity ratios. Gravity is acting in the downward direction $(-y\cos\beta)$ and therefore has components in the streamwise and transverse directions, seen in appearance of two terms with Froude number in Eq.(1).

The velocities satisfy the no-slip boundary conditions at the channel walls:

$$\mathbf{u}_1(y=-n)=0, \quad \mathbf{u}_2(y=1)=0. \tag{2}$$

The disturbed interface $y=\eta(x,z,t)$, which is a surface in the 3D problem, is defined by a unit-length normal vector $\mathbf{n}$, and $\mathbf{t}_1, \mathbf{t}_2$ denote two unit vectors tangential to the interface in the $xy$- and $yz$-planes, respectively:

$$\mathbf{n} = \frac{(-\eta_x, 1, -\eta_z)}{\sqrt{1+\eta_x^2+\eta_z^2}}, \mathbf{t}_1 = \frac{(1, \eta_x, 0)}{\sqrt{1+\eta_x^2}}, \mathbf{t}_2 = \frac{(0, \eta_z, 1)}{\sqrt{1+\eta_z^2}}. \tag{3}$$

Boundary conditions at the interface $y=\eta(x,z,t)$ require continuity of the velocity components and the tangential stresses, and a jump of the normal stress due to the surface tension (in Eqs. (5), (6), and (7), the square brackets denote the jump of the expression value across the interface, e.g., for the jump of quantity $f$, $[f] = f_2 - f_1$)

$$\mathbf{u}_1(y=0) = \mathbf{u}_2(y=0), \tag{4}$$

$$[\mathbf{t}_1 \cdot \mathbf{T} \cdot \mathbf{n}] = \left[\frac{m\mu}{\mu_1}\left(-2\eta_x u_x + (u_y + v_x) - \eta_z(w_x + u_z) + 2\eta_x v_y\right)\right] = 0. \tag{5}$$

$$[\mathbf{t}_2 \cdot \mathbf{T} \cdot \mathbf{n}] = \left[\frac{m\mu}{\mu_1}\left(2\eta_z v_y - \eta_x(w_x + u_z) + (v_z + w_y) - 2\eta_z w_z\right)\right] = 0. \tag{6}$$

$$[\mathbf{n} \cdot \mathbf{T} \cdot \mathbf{n}]$$
$$= \left[p - \frac{m\mu}{\mu_1}\frac{2\mathrm{Re}_2^{-1}}{1+\eta_x^2+\eta_z^2}\left((\eta_x^2-\eta_z^2)u_x + (1-\eta_z^2)v_y - \eta_x(u_y+v_x) - \eta_z(w_y+v_z) + \eta_x\eta_z(u_z+w_x)\right)\right]$$
$$= \mathrm{We}_2^{-1}\frac{\eta_{xx}(1+\eta_z^2)+\eta_{zz}(1+\eta_x^2)-2\eta_x\eta_z\eta_{xz}}{(1+\eta_x^2+\eta_z^2)^{3/2}}. \tag{7}$$

where $We_2 = \rho_2 h_2 u_i^2/\sigma$ is the Weber number, and $\sigma$ is the surface tension coefficient.

The interface displacement and the normal velocity components are coupled by the kinematic boundary condition:

$$v_j = \frac{D\eta}{Dt} = \frac{\partial \eta}{\partial t} + u_j \frac{\partial \eta}{\partial x} + w_j \frac{\partial \eta}{\partial z}. \tag{8}$$

## 3. NON-MODAL GROWTH OF PERTURBATIONS

The perturbed velocity and pressure fields are written as $u_j = U_j + \tilde{u}_j$, $v_j = \tilde{v}_j$, $w_j = \tilde{w}_j$, $p_j = P_j + \tilde{p}_j$, and $\eta = \tilde{\eta}$ for the dimensionless perturbation of the interface displacement. The base flow is steady, laminar, and fully developed. The steady-state velocity $U(y)$ varies only with the cross-section coordinate $y$ (see Appendix). The base flow solution is fully determined by three dimensionless parameters: the viscosity ratio $m$, the flow rate ratio $q = q_1/q_2$, and the inclination parameter $Y = \rho_2(r-1)g\sin\beta/(-dP/dx)_{2S}$. Here $q_j$ is the volumetric feed flow rate of phase $j$ (positive in the x direction), and $(-dP/dx)_{jS} = 12\mu_j q_j/H^3$ is the corresponding superficial pressure drop for single-phase flow in a channel of a height $H = h_1 + h_2$. $U_{jS} = q_j/H$ is the superficial velocity of fluid $j$. Due to consideration of channels of constant height the superficial velocity and the flow rate concepts can be used interchangeably.

To describe the three-dimensional perturbations it is most convenient to add the equation for the wall-normal y-component of vorticity, which is defined as:

$$\omega_j = \frac{\partial u_j}{\partial z} - \frac{\partial w_j}{\partial x}. \tag{9}$$

In the non-modal analysis, we are interested in a short-time behavior of small, but finite amplitude perturbations. Differently from the modal stability analysis, we do not assume an exponential time-dependence of perturbations. At the same time we exploit the homogeneity of the x- and y-directions by assuming solutions of $\tilde{u}_j$, $\tilde{v}_j$, $\tilde{w}_j$, $\tilde{p}_j$, $\tilde{\eta}$, and $\tilde{\omega}_j$ in the form:

$$\tilde{f}(x,y,z,t) = \bar{f}(y,t)e^{i(k_X x + k_Z z)}, \tag{10}$$

where $\bar{f}$ is an amplitude of the corresponding perturbation quantity $\tilde{f}$, $k_X$, $k_Z$ are the dimensionless real wavenumbers in the streamwise and spanwise directions respectively ($k_X = 2\pi h_2/l_X$, $k_Z = 2\pi h_2/l_Z$ with $l_X$ and $l_Z$ being the corresponding wavelengths). In the following discussion the overbars in the notation of the perturbation amplitudes are omitted (e.g., $v_j$ instead of $\bar{v}_j$). Note also that for a more convenient formulation of the Chebyshev collocation method (Barmak et al, 2016a), a new coordinate $y_1 = (y+n)/n$ $(0 \leq y_1 \leq 1)$ is introduced for the part of the channel occupied by the lower phase, while $y_2 = y$ $(0 \leq y_2 \leq 1)$ for the upper phase remains unchanged.

Upon substitution of (10) in the linearized governing equations and boundary conditions, the time evolution of a 3D perturbation is described by the Orr-Sommerfeld equations for the transverse velocity $v$ and the Squire equations for the wall-normal component of vorticity $\omega$ are written in each sublayer:

$0 \leq y_1 \leq 1$
$(-n \leq y \leq 0)$:
$$\frac{\partial}{\partial t}(D_1 v_1) = \left[ik_X\left(-U_1 D_1 + \frac{U_1''}{n^2}\right) + \frac{1}{\text{Re}_1}D_1^2\right]v_1, \tag{11}$$

$0 \leq y_2 \leq 1$
$(0 \leq y \leq 1)$:
$$\frac{\partial}{\partial t}(D_2 v_2) = \left[ik_X\left(-U_2 D_2 + U_2''\right) + \frac{1}{\text{Re}_2}D_2^2\right]v_2, \tag{12}$$

where $D_1 v_1 = \dfrac{v_1''}{n^2} - (k_x^2 + k_z^2) v_1$, $D_1^2 v_1 = \dfrac{v_1^{IV}}{n^4} - 2(k_x^2 + k_z^2) \dfrac{v_1''}{n^2} + (k_x^2 + k_z^2)^2 v_1$,

$D_2 v_2 = v_2'' - (k_x^2 + k_z^2) v_2$, $D_2^2 v_2 = v_2^{IV} - 2(k_x^2 + k_z^2) v_2'' + (k_x^2 + k_z^2)^2 v_2$;

$0 \leq y_1 \leq 1$
$(-n \leq y \leq 0)$:
$$\frac{\partial \omega_1}{\partial t} = -ik_z \frac{U_1'}{n} v_1 + \left[ -ik_x U_1 + \frac{1}{\mathrm{Re}_1} D_1 \right] \omega_1, \tag{13}$$

$0 \leq y_2 \leq 1$
$(0 \leq y \leq 1)$:
$$\frac{\partial \omega_2}{\partial t} = -ik_z U_2' v_2 + \left[ -ik_x U_2 + \frac{1}{\mathrm{Re}_2} D_2 \right] \omega_2. \tag{14}$$

The boundary conditions (b. c.) are as following:

- no-slip at the channel walls:

$y_1 = 0$
$(y = -n)$:
$$v_1 = v_1' = \omega_1 = 0, \tag{15}$$

$y_2 = 1$:
$$v_2 = v_2' = \omega_2 = 0, \tag{16}$$

- continuity of the velocity components (of the perturbed flow) at the interface:

$y_1 = 1, y_2 = 0$:
$$v_1(1) = v_2(0), \tag{17}$$

$y_1 = 1, y_2 = 0$:
$$k_x \left( v_2' - \frac{v_1'}{n} \right) - k_z (\omega_2 - \omega_1) - (k_x^2 + k_z^2)\left( U_2' - \frac{U_1'}{n} \right) \bar{H}_\eta = 0, \tag{18}$$

$y_1 = 1, y_2 = 0$:
$$k_z \left( v_2' - \frac{v_1'}{n} \right) + k_x (\omega_2 - \omega_1) = 0. \tag{19}$$

The (linearized) kinematic boundary condition Eq.(20), the normal stress condition Eq.(21), and continuity of the tangential stresses at the interface Eqs. (22) and (23) reads:

$y_1 = 1, y_2 = 0$:
$$\frac{\partial \bar{H}_\eta}{\partial t} = i\left( v_2(0) - k_x \bar{H}_\eta \right), \tag{20}$$

where $\bar{H}_\eta = i\bar{\eta}$ is introduced in order to avoid complex numbers in the boundary conditions.

$y_1 = 1, y_2 = 0$:
$$\begin{aligned}
&\frac{\partial v_2'}{\partial t} - \frac{r}{n} \frac{\partial v_1'}{\partial t} = ik_x \left[ (v_2 U_2' - v_2' U_2) - \frac{r}{n}(U_1' v_1 - U_1 v_1') \right] \\
&+ \frac{1}{\mathrm{Re}_2} \left[ \left( v_2''' - 3(k_x^2 + k_z^2) v_2' \right) - m\left( \frac{v_1'''}{n^3} - 3(k_x^2 + k_z^2) \frac{v_1'}{n} \right) \right] \\
&- i\bar{H}_\eta (k_x^2 + k_z^2) \left( \frac{\cos\beta}{\mathrm{Fr}_2}(r-1) + \frac{(k_x^2 + k_z^2)}{\mathrm{We}_2} \right),
\end{aligned} \tag{21}$$

$y_1 = 1, y_2 = 0$:
$$k_z \left[ \left( v_2'' + (k_x^2 + k_z^2) v_2 \right) - m\left( \frac{v_1''}{n^2} + (k_x^2 + k_z^2) v_1 \right) \right] + k_x \left( \omega_2' - m\frac{\omega_1'}{n} \right) = 0, \tag{22}$$

$y_1 = 1, y_2 = 0$:
$$\begin{aligned}
&k_x \left[ \left( v_2'' + (k_x^2 + k_z^2) v_2 \right) - m\left( \frac{v_1''}{n^2} + (k_x^2 + k_z^2) v_1 \right) \right] \\
&- k_z \left( \omega_2' - m\frac{\omega_1'}{n} \right) - (k_x^2 + k_z^2)\left( U_2'' - m\frac{U_1''}{n^2} \right) \bar{H}_\eta = 0.
\end{aligned} \tag{23}$$

Further analysis is provided for a numerical solution of the above time-dependent PDE problem defined in the y-direction and in time. The Chebyshev collocation method used in our previous studies (Barmak, 2016a,

b) is applied. The solution vector $\mathbf{q} = \begin{pmatrix} v_j & \omega_j & \bar{H}_\eta \end{pmatrix}^T$ is approximated in each sublayer as a truncated series of $N$ shifted Chebyshev polynomials ($T_i(y)$, $0 \leq y \leq 1$) with time-dependent coefficients:

$$q^{(j)} = \sum_{i=1}^{N} d_i^{(j)}(t) T_i(y), \tag{24}$$

The collocation points are the roots of the N-th order Chebyshev polynomial. Upon discretization in the y direction, the problem reduces to the following dynamical system:

$$\mathbf{B_1} \frac{d\mathbf{q}}{dt} = \mathbf{A_1} \mathbf{q}, \tag{25}$$

where $\mathbf{A} = \mathbf{B_1}^{-1} \mathbf{A_1}$ is the linearized dynamical matrix operator. The formal solution of the initial value problem (25) can be written as:

$$\mathbf{q}(t) = e^{\mathbf{A}t} \mathbf{q}(0). \tag{26}$$

At the first stage, the spectrum of the eigenvalues and eigenvectors of matrix $\mathbf{A}$ should be found. To this aim, the following generalized eigenvalue problem has to be solved:

$$\lambda \mathbf{d} = \mathbf{A} \mathbf{d}, \tag{27}$$

where the order of the eigenvalue problem is $4N+1$. It is done by the QR eigensolver, as in our previous studies (Barmak, 2016a, b). Note that the eigenvalue $\lambda = \lambda_R + i\lambda_I$ obtained from Eq. (27) corresponds to the complex time increment of the corresponding modal stability problem, and $\lambda_R$ determines the growth rate.

The non-modal analysis is performed following the approach of Reddy and Henningson (1993), which is briefly described below. The solution of Eq. (26) can be rewritten in terms of the $L$ leading eigenvalues (i.e., those with the largest real part) of the evolution matrix $\mathbf{A}$ and their corresponding eigenvectors:

$$\mathbf{q} = \sum_{l=1}^{L} \mathbf{p}_l e^{\lambda_l t} k_l(0), \tag{28}$$

where $\mathbf{p}_l$ is the $l$-th eigenvector of matrix $\mathbf{A}$ ($l$-th column of the eigenvectors matrix $\mathbf{P}$) and $k_l(0)$ is the $l$-th component of the vector of coefficients $\mathbf{k}(0) = \mathbf{P}^{-1} \mathbf{q}(0)$.

A physically relevant quantity for measuring transient growth of perturbations caused by a non-orthogonality of the eigenvectors of matrix $\mathbf{A}$ is the energy norm (which can be alternatively written as a special form of the inner product of perturbation vectors, denoted as $\langle \rangle_E$ below) that in addition to the kinetic and interfacial components (e.g., South and Hooper, 1999) includes also a gravitational component:

$$\|\mathbf{q}\|_E^2 = \langle \mathbf{q}^*, \mathbf{q} \rangle_E = E = E_{kin} + E_{int} + E_g = \frac{1}{2(k_X^2 + k_Z^2)} \left[ r \int_0^1 \left( \frac{|v_1'|^2}{n^2} + (k_X^2 + k_Z^2)|v_1|^2 + |\omega_1|^2 \right) dy_1 \right.$$
$$\left. + \int_0^1 \left( |v_2'|^2 + (k_X^2 + k_Z^2)|v_2|^2 + |\omega_2|^2 \right) dy_2 \right] + \frac{(k_X^2 + k_Z^2)|\bar{H}_\eta|^2}{2We_2(n+1)} + \frac{(r-1)\cos\beta|\bar{H}_\eta|^2}{4\operatorname{Fr}_2(n+1)}, \quad (29)$$

where the superscript * refers to the complex-conjugate.

Reddy and Henningson (1993) showed that the energy norm can be transformed into an equivalent 2-norm that can be determined using the singular value decomposition (SVD) (or, alternatively, by Cholesky decomposition) of a positive defined Gram matrix $\mathbf{S}$, whose elements are inner products of the eigenvectors of matrix $\mathbf{A}$, i.e., $\mathbf{S} = \langle \mathbf{p}^*, \mathbf{p} \rangle_E = \mathbf{U}\boldsymbol{\Sigma}\mathbf{V}^H = \mathbf{U}(\boldsymbol{\Sigma})^{1/2}(\boldsymbol{\Sigma})^{1/2}\mathbf{U}^H = \mathbf{F}^H\mathbf{F}$. Here the matrices $\mathbf{U}$ and $\mathbf{V}$ are unitary matrices and matrix $\boldsymbol{\Sigma}$ is a diagonal matrix, with the singular values of $\mathbf{S}$ on its diagonal. $\mathbf{F}$ is a lower triangular matrix with strictly positive diagonal entries. The superscript $^H$ refers to the Hermitian transpose of the corresponding matrix. Then:

$$\|\mathbf{q}\|_E^2 = \mathbf{k}^* \langle \mathbf{p}^*, \mathbf{p} \rangle_E \mathbf{k} = \mathbf{k}^*\mathbf{S}\mathbf{k} = \mathbf{k}^*\mathbf{F}^H\mathbf{F}\mathbf{k} = \|\mathbf{F}\mathbf{k}\|_2^2. \quad (30)$$

The energy growth function is then given by (see Schmid and Henningson (2000) for details):

$$G(t) = \max_{E(0) \neq 0} \frac{E(t)}{E(0)} = \max_{\mathbf{q}(0) \neq 0} \frac{\|\mathbf{q}(t)\|_E^2}{\|\mathbf{q}(0)\|_E^2} = \max_{\mathbf{q}(0) \neq 0} \|\mathbf{F}e^{\mathbf{A}t}\mathbf{F}^{-1}\|_2^2 = \|\mathbf{M}(t)\|_2^2 = \sigma_1^2(t). \quad (31)$$

where $\sigma_1(t)$ is the first (maximal) singular value of the matrix $\mathbf{M}(t) = \mathbf{F}e^{\mathbf{A}t}\mathbf{F}^{-1}$. Thus, the computation of the growth function is reduced to the calculation of the singular value decomposition (SVD) of the propagator operator $\mathbf{M}$ ($\mathbf{M} = \mathbf{U}_\mathbf{M}\boldsymbol{\Sigma}_\mathbf{M}\mathbf{V}_\mathbf{M}^H$). As discussed in Schmid and Henningson (2001), non-orthogonality of the eigenvectors of the matrix $\mathbf{A}$ can lead to transient energy growth of perturbation, so that even in linearly stable regimes the growth function $G(t)$ can exceed unity inside certain time interval. At large times it decays or grows exponentially ($\lambda_R < 0$ for all eigenvalues in Eq. (27) or $\lambda_R > 0$ for some of eigenvalues, respectively). The optimal initial conditions, that yield the maximal energy gain (i.e., the maximal value of $G(t)$, $G_{MAX} = G(t_{opt}) = \sigma_1^2(t_{opt})$), can be defined by SVD:

$$\mathbf{q}_0 = \mathbf{P}\mathbf{F}^{-1}\mathbf{v}_{\mathbf{M}1}, \quad (32)$$

where $\mathbf{v}_{\mathbf{M}1}$ is the first right-singular vector of matrix $\mathbf{M}(t_{opt})$ and $t_{opt}$ is time at which $G_{MAX}$ is attained.

It is important to note that the non-modal growth is not calculated in the full space of matrix $\mathbf{A}$, but in its subspace defined by the $L$ leading eigenvectors. $L$ is set to be large enough to achieve convergence of the growth function. Compared to the linear (modal) stability analysis (Barmak et al., 2016a, b), the non-modal growth study is more computationally demanding, since many ($L \gg 1$) leading eigenmodes must be calculated to obtain converged growth function. Most of the results presented below were computed using the truncation number $N = 80$, which results (for 2 layers and 2 variables) in 320 degrees of freedom in the numerical model. It was found that the $L = 80$ least stable modes assure the convergence of the numerical solution as demonstrated in Table I. Calculations were performed using double-precision floating-point numbers (i.e., each number occupies 64 bits of memory), which were found to give the same results as calculations with higher (quadruple) precision. The numerical solution was verified by comparison with the solution of Yecko (2008) for zero-gravity systems (the details are given below).

TABLE I. Convergence of the growth function with increasing the truncation number $N$ and number of least stable eigenmodes $L$.

|  | Yecko (2008): $r = 1.11, m = 0.5,$ $h = 0.5, \text{Re}_2 = 405, \text{We}_2 = 0.9$ $k_X = 1, k_Z = 0$ | | Horizontal air-water flow ($H$=0.02m, point **A** in Fig. 12): $r = 1000, m = 55,$ $h = 0.5, \text{Re}_2 = 49.5, \text{We}_2 = 0.0012$ $k_X = 0, k_Z = 3$ | |
|---|---|---|---|---|
|  | $G_{MAX}$ | $H_{\eta\_MAX}$ | $G_{MAX}$ | $H_{\eta\_MAX}$ |
| $N = 50, L = 50$ | 2.6631580 | 2.00998456 | 101.67993 | 1.07910338 |
| $N = 50, L = 80$ | 2.6639072 | 2.00745977 | 101.68118 | 1.07664110 |
| $N = 80, L = 80$ | 2.6639084 | 2.00745729 | 101.68104 | 1.07732035 |
| $N = 100, L = 100$ | 2.6639959 | 2.00709943 | 101.68109 | 1.07712025 |
| $N = 120, L = 120$ | 2.6640392 | 2.00690639 | 101.68111 | 1.07702110 |

\* $G_{MAX}$ is $\max(G(t))$, $H_{\eta\_MAX}$ is maximal growth of the interfacial displacement amplitude $H_\eta = |\bar{H}_\eta(t)|/|\bar{H}_\eta(t=0)|$.

## 4. RESULTS AND DISCUSSION

The non-modal growth of both 2D (in the plane of the flow) and 3D perturbations are studied in zero-gravity conditions as well as in the gravitational field. Due to the large number of parameters involved in the non-modal analysis, the present study is focused on the several characteristic systems considered in our previous works (Barmak, 2016a, b). The transient energy growth can be particularly important in two-phase flows if it is associated with significant fluid-fluid interface deformations. In such cases, transition (possibly in a nonlinear regime) from stratified flow to other flow patterns may take place. Thus, we analyze the results on the time evolution of the interface displacement amplitude of optimal perturbations (i.e., perturbations associated with $G_{MAX} = \max(G(t))$. The time evolution of the interface is represented by the maximal (in space) displacement of the interface at a particular instant, $H_\eta = |\bar{H}_\eta(t)|/|\bar{H}_\eta(t=0)|$, and $H_{\eta\_MAX}$ denotes its maximal value. It is worth noting that in linear analysis all perturbation amplitudes are defined up to some arbitrary defined constant (that is considered to be small), a relevant reference value should be taken. In the present study, an initial value of $|v|$ at the interface has been taken as the reference value for all perturbation amplitudes.

Since one of the goals of this work is to study the significance of transient growth for the prediction of the smooth-stratified flow boundaries, linearly stable (subcritical) conditions are examined. We seek for flow parameters that are associated with $H_{\eta\_MAX} > 1$ and accompanied by transient energy growth, which may be a precursor to flow pattern transition in the subcritical regime. Note that in the framework of the non-modal analysis a precise threshold for the transient growth at which the non-linear mechanisms become non-negligible cannot be predicted. In the linearly unstable regime, the interface deformation grows exponentially in time and leads to transition to other flow patterns. Therefore, such conditions are of limited interest in this study.

### A. Flows under zero-gravity conditions

In previous works on the non-modal growth of perturbations in stratified two-phase flows (van Noorden et al., 1998; South and Hooper, 1999; Yecko, 2008; Ó Náraigh et al., 2014), only systems under zero-gravity conditions (i.e., $g = 0$, or fluids of similar densities, $r \approx 1$) were considered. As an example for validation of our numerical calculations and for the purpose of the further discussion of new findings, we start with presenting the results of our analysis for the case study considered by Yecko (2008), labeled as "YC" below. This case is characterized by equal layer thicknesses of two fluids of similar densities under zero-gravity conditions, and is unambiguously defined (except for surface tension, which requires a value of We) by the following dimensionless parameters (subscript 'Y' denotes a parameter value corresponding to the Yecko's notation): $h = 0.5 \left( n_Y = 1 \right)$, $r = 1.11$ (or $r_Y = 0.9$), $m = 0.5$ $\left( m_Y = 2 \right)$, $\text{Re}_2 = 405 \left( \text{Re}_Y = 900 \right)$. The results of the spectra were found to be identical to those of Yecko (2008), e.g., Fig. 2. $N = 80$ collocation points were found to be sufficient for convergence of the spectrum.

As shown in Fig. 2, when surface tension effects are prominent $\left( \text{We}_2 = 0.9 \right)$, the flow is stable with respect to the considered 2D perturbation as all $\lambda_R$ $\left( = \text{Re}(\lambda) \right)$ are negative. At larger $\text{We}_2 = 90$ (reduced surface tension) the flow is unstable and its leading eigenvalue is $\lambda = 2.005797 \cdot 10^{-3} - i \cdot 1.0269497$ (compared to $s_Y = i\lambda = 1.026949 + i \cdot 2.006 \cdot 10^{-3}$ in Yecko, 2008). The results for the transient energy growth as a function of dimensionless time, $G(t)$, presented in Fig. 3(a)) for different values of $\text{We}_2$, are identical to those presented in Figure 4(a) in Yecko (2008). Obviously, in all zero-gravity cases the energy norm defined in Eq. (29) accounts only for the kinetic and interfacial components. As shown in the figure, the reduction of surface tension (increase of $\text{We}_2$) results in larger maximal energy growth, $G_{MAX}$ (Fig. 3(a)), and is accompanied by destabilization of the flow for $\text{We}_2 = 90$. In the latter case, an exponential (modal) growth is obtained at larger times (not shown in the figure).

In contrast to the effect of surface tension on $G_{MAX}$, its effect on the evolution of the interface displacement amplitude of the associated optimal perturbation, $H_\eta$, is not so clear (Fig. 3(b)). For all the cases considered, initial growth of the interface displacement amplitude is observed, while the peak for $\text{We}_2 = 9$ is the highest.

Figure 4 shows the evolution with time of the kinetic and interfacial energies contribution to the total energy gain of the optimal perturbation (all the components of energy are normalized by a value of the total energy at $t = 0$). The contribution of the kinetic energy is noticeably the largest. At the same time, the interfacial energy addition, while never exceeding 30% of the total energy, cannot be neglected. Moreover, as it was already noticed by Yecko (2008), the addition of the interfacial energy makes the total energy growth function smoother (e.g., see Fig. 4(a)). It exhibits a non-monotonic behavior and its maximum corresponds to the maximum of the kinetic energy, while the large interface deformations occur at other times (Fig 3(b)). Although the interface displacement growth is larger for lower surface tension $(\text{We}_2 = 9)$, the relative contribution of the associated interfacial energy is smaller (Fig. 4(b)). Nevertheless, it is important to mention that the inclusion of the interfacial energy in the energy norm is essential for correct calculation (and convergence) of the transient growth and for the identification of the optimal perturbation.

It is well-known that the Squire theorem does not hold for the non-modal analysis, and that in shear flows 3D perturbations usually gain larger energy than 2D ones (e.g., Gustavsson, 1991; Vitoshkin et al., 2012). The maximal energy growth, $G_{MAX}$ as a function of the streamwise and spanwise wavenumbers for YC is shown in Fig. 5. The spanwise perturbation $(k_X = 0, k_Z \approx 3)$ is found to attain the largest energy gain (the global maximum of the energy growth, $G_{MAX}^{MAX}$, i.e. the maximum of $G_{MAX}$ over all $k_X$ and $k_Z$, is located on the $k_Z$-axis). The energy of the optimal perturbation in this case grows by a factor of 70, which is much higher than for the 2D streamwise perturbations considered previously.

Growth functions for some specific spanwise and oblique perturbations are shown in Fig. 6 (a), all of them attain larger energy gain than that of the 2D streamwise perturbation considered above. However, when examining the evolution of the interface displacement amplitude of the optimal perturbations, the 2D streamwise $(k_X = 1, k_Z = 0)$ perturbation yields larger growth of the interface displacement than the spanwise (with $k_X = 0$) perturbations. The maximal growth of the interface displacement among several considered wavenumbers is obtained by an oblique optimal perturbation with $k_X = 1$ and $k_Z = 3$ ($H_{\eta\_MAX} \approx 6$, Fig. 6(b)). The oscillatory behavior of the interface displacement growth (Fig. 6(b)) is a consequence of the presence of eigenmodes with non-zero frequencies (i.e., $\text{Im}(\lambda)$ values) that contribute to the transient growth. It can be concluded that in agreement with the claim of Yecko and Zaleski (2005) and Yecko (2008), the three-dimensional spanwise perturbation (i.e., $k_X = 0$, $k_Z = 3$) is found to exhibit the strongest non-modal behavior. However, they do not involve growth of the interface displacement. Thus, the transient growth of these perturbations may trigger non-linear mechanisms within one of the layers, but not a change of the structure of the interface (i.e., the flow pattern). Nevertheless, in this particular case, there are other perturbations (e.g., $k_X = 1$, $k_Z = 3$, which is not the one that gives $G_{MAX}^{MAX}$) that exhibit growth of the interface displacement. However, once all wavenumber perturbations are considered, the modal analysis should also be applied to find out whether the stratified-smooth flow is linearly stable.

Applying the modal stability analysis (e.g., Barmak et al., 2016) we found that for flow parameters considered above (YC) the smooth-stratified flow is actually linearly unstable. As shown in Fig. 7(a) even for the largest surface tension (i.e., the smallest Weber number, $\text{We}_2 = 0.9$), there is a range of 2D perturbations around the most amplified mode $(k_{X\_\max} = 0.1)$ that grow exponentially with time. The growth rate obtained in the modal stability analysis is maximal for 2D streamwise perturbations (in agreement with the Squire's theorem, see details in Barmak et al., 2017). Fig. 7(b) maps the modal growth rate of 2D and 3D perturbation, i.e. whether perturbations exponentially grow or decay in time ($\lambda_R > 0$ or $\lambda_R < 0$, respectively). As shown, the global optimal perturbation ($k_X = 0$, $k_Z = 3$) is a (linearly) stable mode. Therefore, although it initially grows, it will decay at longer times. Considering the fact that in the unstable regime there are already many 2D and 3D modes that grow exponentially with time (including the interface displacement amplitude), and thereby lead to flow pattern transition, the non-modal analysis for such conditions is of limited interest in this work. Therefore, in the following we examine only linearly stable cases.

In contrast to the considered above flow, air-water flow is characterized by large viscosity and density ratios ($m = 55$ and $r = 1000$, respectively). Under zero-gravity conditions, such a flow is found to be stable only when the water is faster than the air (i.e., above the critical holdup, $h > h_{cr} = \sqrt{m}/(1+\sqrt{m}) = 0.88$, or the critical flow rate ratio line, $q_{cr} = \sqrt{m} = 7.416$, in Fig. 8), and for relatively low flow rates of the phases (see details in Barmak et al., 2016a). In Fig. 8, the stability map is plotted in the (dimensional) superficial velocity coordinates, for the channel of 2 cm height. The air density and dynamic viscosity are $\rho_2 = 1\,\text{kg}/\text{m}^3$ and $\mu_2 = 1.8181 \cdot 10^{-5}\,\text{Pa}\cdot\text{s}$, respectively, and the air-water surface tension coefficient is $\sigma = 0.072\,\text{N}/\text{m}$. The solid (blue) curve confines the region of operational conditions which was found to be (linearly) stable with respect to all perturbations in the framework of modal stability analysis. Each point along the stability boundary corresponds to water and air superficial velocities for which the flow is neutrally stable for a particular 2D perturbation (denoted as the critical perturbation) and is stable with respect to all other perturbations. For higher superficial velocities, beyond the neutral stability boundary, there is a range of wave numbers for which the perturbations are amplified. For the purpose of the non-modal analysis, point **A** $\left(h = 0.95, U_{1S} = 0.05\,\text{m/s}, U_{2S} = 0.002\,\text{m/s};\ \text{Re}_2 = 49.5; \text{Fr}_2 = 8.26 \cdot 10^{-2}; \text{We}_2 = 1.12 \cdot 10^{-3}\right)$ located within the stable region is selected.

Examination of the maximal energy growth of various perturbations at point **A** (Fig. 9(a)) shows that long-wavelength streamwise perturbations $(k_X \to 0, k_Z = 0)$ are the optimal ones with an energy increase of up to 250 times compared to the initial state. When the transient energy growth attains several orders of magnitude (as in the considered case), it is reasonable to assume that further flow evolution will be affected by non-linear effects. The growth function vs. the dimensional time obtained for several wave numbers is presented in Fig. 9(b). As expected, after the initial energy gain, the growth function subsequently decays at larger times, owing to the linear stability of the flow. In order to assess the dimensional time by which the maximal non-modal growth is attained, characteristic "residence time" $\tau$ can be referred to. To this aim, the characteristic residence time is defined as $\tau = 500H/u_i$ to signify whether the transient growth of the perturbations can be observed only in very long channels (e.g., with length $> 500H$), or in lab-scale (short) channels as well. The results show that the maximal growth rate can be attained in such a channel within the flow residence time. However, although the total energy grows for all the considered perturbations, none of them exhibit growth of the interface displacement. The interface displacement amplitude was found to decrease monotonically or oscillatory in time (therefore, it is not shown). On the other hand, growth of the interface displacement has been found for some perturbations at point **B,** which situated in the thin strip of the stable region just above the critical holdup line in Fig. 8 $\left(h = 0.89, U_{1S} = 0.09\,[m/s], U_{2S} = 0.01\,[m/s];\ \text{Re}_2 = 17.4; \text{Fr}_2 = 1.08; \text{We}_2 = 6.58 \cdot 10^{-4}\right)$. For example, Fig. 10(c) shows that $H_\eta$ increases by a factor of 8 for the optimal perturbation with $k_X = 0.1, k_Z = 0.3$. The maximal transient energy growth is also higher at this point (about 750, Fig. 10(a), (b)) and corresponds to a spanwise perturbation $(k_X = 0, k_Z = 0.3)$. These results imply that a smooth-stratified flow may not exist in that narrow (subcritical) region due to non-modal instability.

### B. Horizontal flows

Gravity is known to change dramatically the stability boundaries of stratified two-phase flow (see, e.g., Barmak et al., 2016a). However, its effect on non-modal stability has not been investigated in the literature. In order to investigate the effect of gravity we revisit YC, but considering the flow in the gravitational field. For such conditions, the flow is linearly stable (contrarily to the corresponding zero-gravity case). In this case, and throughout the following test cases involving the gravity effect, the energy norm accounts also for the gravitational component. Despite the presence of the gravity $(Fr_2 = 8.26 \cdot 10^{-2})$, the results on the transient energy growth (Fig. 11(a)) are almost identical to those of the corresponding zero-gravity case (Fig. 6(a)). Such results can be attributed to the small density difference and, consequently, a negligible role of buoyancy in this two-phase system. Nevertheless, the gravity affects the interface displacement amplitude of the optimal perturbation (Fig. 11(b)), whose growth is smaller than in the corresponding zero-gravity system (e.g., for $k_X = 1, k_Z = 3$  $H_{\eta\_MAX} \approx 3.6$ in this case versus $H_{\eta\_MAX} \approx 6$ in the zero-gravity case, Fig. 6(b)).

Air-water flow is an example of a system characterized by a large density ratio $(r = 1000)$ as well as large viscosity ratio $(m = 55)$. The stability map found in the framework of modal stability analysis is presented in Fig. 12. It can be seen that indeed the stability map changes drastically in the presence of gravity. Under normal gravity, the flow is stable in a wider range of superficial velocities (cf. Figures 8 and 12). In the non-modal analysis, flow conditions corresponding to points within the stable (subcritical) region (not far from the stability boundary) are investigated in order to examine a possibility of the non-modal growth of the interface displacement that may be a precursor to flow pattern transition. Point **A** on the stability diagram corresponds to a base flow with equal layers thickness $(h = 0.5, U_{2S} = 0.232\,\text{m/s}, U_{1S} = 0.026\,\text{m/s};\ Re_2 = 49.5;$ $Fr_2 = 8.26 \cdot 10^{-2};\ We_2 = 1.12 \cdot 10^{-3})$. The maximal transient energy growth for 3D perturbations of various $k_X$ and $k_Z$ (Fig. 13) is lower than that obtained for air-water flow under zero-gravity (≈100 versus ≈600 for point **B** of air-water zero-gravity, Fig. 10(b)). Here too, $G_{MAX}^{MAX}$ is attained by the spanwise perturbation, $k_X = 0,\ k_Z = 3$. The 2D streamwise perturbations show relatively insignificant growth. The results of the growth function vs. time for several wavenumbers (Fig. 14(a)) indicate that the maximal values of energy growth are reached for relatively short times (<10s, in comparison to the residence time, $\tau$, of ≈111s). Not only spanwise, but also oblique perturbations attain substantial energy growth, especially with wavenumbers close the one corresponding to $G_{MAX}^{MAX}$.

The interface displacement amplitude of the optimal perturbation associated with $G_{MAX}^{MAX}$ (Fig. 14(b)) does not show any initial growth and monotonically decays in time. However, the figure demonstrates the evolution of the interface displacement caused by several other selected optimal perturbations. Growth of the interface displacement amplitude (ten times and higher) is observed for some oblique perturbations $(k_X = k_Z = 1$ and $k_X = 1, k_Z = 3)$. Even though the initial interface deformation is considered to be of small amplitude (as assumed in the framework of linear analysis), such growth of the interfacial displacement, along with the transient energy growth of perturbations, may trigger nonlinear mechanisms of instability. Consequently, the interface may not return to its initial smooth structure, and flow pattern transition (to stratified-wavy flow pattern) may take place within the region corresponding to subcritical conditions .

Fig. 15 shows the profiles of the absolute value of the transverse velocity amplitude $|v|$ and of the y-vorticity $|\omega|$ of the optimal perturbation $(k_X = 0, k_Z = 3)$ associated with $G_{MAX}^{MAX}$ at $t = 0$ and 5.2s (when $G(t) = G_{MAX}^{MAX}$). During the whole evolution process, the perturbation velocity profile has two maxima, one in each of the layers, and the larger one is within the lower (heavy) layer. The transverse velocity decays in time, while the energy growth is associated with the growth of the y-vorticity. The maximal value of vorticity is reached in the bulk of the heavy layer, which in this case is more viscous. Such perturbation profiles can be attributed to much higher density of water, while for systems with fluids of similar densities (e.g., Yecko, 2008), the maximum of the optimal perturbation profile is observed in the less viscous layer. It should be noted that the characterization of the perturbation based on the location of the maximum in the perturbation profile (e.g., shear mode or interfacial mode) cannot be obtained in a similar manner to that used in the modal stability analysis (see discussion in Barmak et al. 2016a, b), since the perturbation evolution in time is different for each point in the flow cross-section.

The optimal perturbation structure is further examined by considering the velocity vectors distribution and streamwise velocity contours on the *yz*-plane. Two snapshots at $t = 0$ and 5.2s are shown in Fig. 16(a) and (b) respectively. Two rows of streamwise vortices are observed in the initial state, the centers of the stronger ones are in the bulk of the lower layer. The weaker vortices in the upper layer are counter-rotating and in phase with respect to their counterparts in the lower layer. By $t = 5.2$ s (maximal energy growth), the vortices in the *yz*-plane are getting much weaker than at the beginning (implying the decrease of $v$ and $w$) and the energy growth is associated with the growth of the perturbation of streamwise velocity in the lower layer (accompanied by the growth of the y-vorticity $|\omega|$), while the interface amplitude is decaying. It should be pointed out that this perturbation, as well as those of all other spanwise perturbations (i.e., $k_X = 0$), is a standing wave. This is due to the fact that a spanwise non-modal perturbation is a result of superposition of eigenmodes (Eq. (28)), where two of them are counter-propagating with the same phase speed, while all the other are standing waves (i.e., $\text{Im}(\lambda) = 0$). On the other hand, streamwise and oblique perturbations are traveling waves in the direction of the wavevector **k,** and their phase speed is time-dependent.

Another interesting test case is horizontal air-water flow in a larger (20cm-height) channel. For this case, the modal stability analysis predicts that low air flow rates stabilize the corresponding single-phase water flow, whereby the critical water superficial velocity for destabilization of the flow is much higher than that predicted by the single-phase flow laminar limits (see Barmak et al., 2016a). The superficial phase velocity corresponding to the single-phase flow laminar limit (dashed lines for the gas and liquid phases in Fig.17) is determined by the critical Reynolds number, $\text{Re}_{Cr} = 5772$ (Orszag, 1971). For the non-modal analysis, subcritical point **A** $(h = 0.83; U_{2S} = 0.1 \text{m/s}; U_{1S} = 0.4 \text{m/s}; \text{Re}_2 = 1332.1; \text{Fr}_2 = 1.66; \text{We}_2 = 0.25)$ near the (modal) stability boundary is selected (Fig. 17). Similarly to the previous test cases, a spanwise perturbation (with $k_X = 0$) is found to exhibit the maximal energy growth, $G_{MAX}^{MAX}$. Moreover, the value of $G_{MAX}^{MAX}$ is much higher than those found in the previous cases (of the order of $10^6$, Fig. 18(a), (b)). The period of initial growth is longer than in the other cases as well, and largely exceeds the residence time. Thus, smaller but still substantial transient energy growth can be expected in short channels that can trigger non-linear mechanisms leading to the flow destabilization. Such strong amplification can be attributed to the dominance of the (destabilizing) inertial forces over the (stabilizing) viscous forces in the thick water layer characterized by a relatively high water Reynolds number ($\text{Re}_{1S} = \rho_1 U_{1S} H / \mu_1 = 80,000$, i.e. far beyond the single phase $\text{Re}_{Cr}$). This leads to very large growth of the kinetic energy of perturbations. A similar picture of the transient energy growth is observed for most of this region above the single phase water flow laminar flow limit (blue dashed line in Fig.17). Although exhibiting a strong non-modal behavior in the total energy, the optimal perturbations corresponding to $G_{MAX}^{MAX}$ are not associated with growth of the interface displacement. In this flow-rate region, the energy growth is dominated by the increase of the perturbation kinetic energy and is expected to lead to flow transition (similarly to single-phase shear flows) in the thick water layer, although the modal analysis predicts a stable laminar flow. Nevertheless, it is important to note that oblique wavenumbers can result in growth of the interfacial displacement amplitude also in this case, which may block the narrow air-flow passage (due to the large water holdup, $h=0.83$, thereby leading to plug/slug flow). These results are not reported here, because for those wavenumbers we were not able to arrive to a convergent solution even by using as many as 300 collocation points in each layer.

While the part of the linearly stable region on the flow pattern map above the water laminar limit was found to be a subject for strong non-modal growth, it is of interest to check also the conditions with lower subcritical water superficial velocities, e.g., point **B** in Fig. 17 $\left(h=0.49, U_{2S}=0.2\,\text{m/s}, U_{1S}=0.02\,\text{m/s};\right.$ $\left.\text{Re}_2=397.5; \text{Fr}_2=4.93\cdot10^{-3}; \text{We}_2=7.07\cdot10^{-3}\right)$ As expected, the transient energy growth (Fig. 19(a)) is much smaller than at point **A**, however, it is still large in comparison with similar conditions in the smaller channel of 2-cm height. A spanwise perturbation with $k_Z=3$ encounters the maximal possible energy growth at this point. Growth of the interfacial displacement amplitude has been found for a particular oblique perturbation $\left(k_X=0.5, k_Z=1\right.$, Fig. 19(b)), that also encounters some relatively mild growth in energy. This case is also characterized by slow growth in time. However, due to relatively long residence time $\left(t(G_{MAX})<\tau\right)$, large growth can be attained even in short channels.

The stable stratified flow region is known to shrink when operating in mini-and micro-channels. This has been also predicted by the modal stability analysis (see Barmak et al., 2016a for a mini-channel of 2-mm height). The stability of the flow in the region of low air and high water superficial velocities was found to be dominated by surface tension and resembles zero-gravity conditions. Point **A** $\left(h=0.95; U_{1S}=0.4\,\text{m/s};\right.$ $U_{2S}=0.015\,\text{m/s}; \text{Re}_2=3.11; \text{Fr}_2=365.04; \text{We}_2=4.61\cdot10^{-4}\right)$ in this region has been considered for the non-modal analysis (Fig. 20). The optimal perturbation that encounters the maximal transient energy growth $\left(G_{MAX}^{MAX}\approx100,\text{ see Fig. 21(a)}\right)$ is spanwise $\left(k_X=0, k_Z=0.1\right)$. In addition, substantial growth of the interface displacement amplitude was found for the oblique optimal perturbations with $k_X=0.1$ and $k_Z=0.1$ (Fig.

21(b)). For such a high holdup of water $(h = 0.95)$ even moderate growth of the interfacial displacement amplitude may block the air flow passage and lead to a transition to other flow patterns (probably to bubbly or plug flow). In fact, the non-modal analysis shows that the whole region of high water holdups is subject to large growth of the interface displacement amplitude, implying possibility of reduction of the stable region of stratified-smooth flow. On the other hand, for larger air superficial velocities and thinner water layers (below the critical holdup line) the transient energy growth is rather small. In addition, no growth of the interface displacement of the optimal perturbations has been found in that part of the stable region.

### C. Inclined flows

The stability map for air-water flow in $\beta = 1°$ downward inclined channels obtained by the modal analysis is shown in Fig. 22. The concurrent downflow was found to be stable only in the region of sufficient low water superficial velocities. The stable region is limited to significantly lower liquid flow rates (and holdups) compared to horizontal channel (see Barmak et al., 2016a) and is in agreement with experimental observations in downward inclined pipe flow (Barnea et al., 1982). Therefore, it can be expected that non-modal energy growth does not affect the transition to other flow patterns. The subcritical point **A** in Fig. 22, which corresponds to relatively high air flow rates and is located near the stability boundary, is selected to examine non-modal energy growth $(h = 0.03; U_{1S} = 0.001 \,\text{m/s}; U_{2S} = 2 \,\text{m/s}; \text{Re}_2 = 49.5; \text{Fr}_2 = 1.14 \cdot 10^{-2};$ $\text{We}_2 = 5.83 \cdot 10^{-4})$. As can be seen from the contours of the maximal energy growth (Fig. 23(a)), the energy growth is quite significant for these conditions. However, as shown in in Fig. 23(b), the period of energy growth is very short in time and lasts about a second (or even less). In fact, in this case of low water holdup, the transient growth is strongly affected by the growth of the kinetic energy in the bulk of the thick air layer. Therefore, the relevant time scale for referring to the transient growth, should be the air phase residence time (based on the characteristic air velocity, which is much higher than the interfacial velocity), $\tau_2 = 500H/u_{2S} = 5\,\text{s}$. Indeed, this time scale is of the order of the duration of the transient perturbation growth.

The interface displacement of most of the optimal perturbations tends to decay, and only some oblique perturbations show slight short-time growth (Fig. 24 (a)). These results suggest that such perturbations are not expected to impact the stability of the stratified flow configuration predicted by the modal linear analysis. The optimal perturbations can trigger non-linear mechanisms within the dominating (air) layer, but do not result in significant interfacial deformations. It is worth emphasizing that in downward inclined channels the liquid flow is driven by gravity, and it is essential to consider the potential gravitational energy in the definition of the energy norm. This is demonstrated in Fig. 24(b), where results of the transient energy growth obtained by considering different energy norms are shown. The difference between complete and incomplete energy norm definitions is prominent, indicating that only the full energy norm should be used for obtaining meaningful results. It is worth noting that the optimal perturbations associated with those norms are also different.

The optimal perturbation structure is further examined by considering the velocity vectors distribution and streamwise velocity contours on the *yz*-plane. Two snapshots, at $t = 0$ and 0.4s (maximal energy growth), are shown in Fig. 25(a) and (b), respectively. A row of streamwise vortices with the centers in the bulk of the air layer is observed. By 0.4s, the vortices are getting much weaker and the energy growth is associated with the growth of the streamwise velocity perturbation in the lower (heavy) phase, while the interface amplitude is decaying with oscillations.

In slightly upward inclined flows ($\beta = 0.1°$), there is a range of superficial velocities where three possible stratified flow configurations with different holdups exist for specified superficial velocities. The triple solution region is characterized by sufficiently high air flow rates and low water flow rates, which enable also a flow configuration where the entire water layer is dragged upward and backflow of the liquid near the bottom wall is avoided. This flow configuration corresponds to the lower holdup solution, while in the two additional solutions of higher liquid holdups (middle and upper holdup solutions), backflow of the liquid is typically obtained. Linear (modal) analysis revealed that the lower and middle holdup solutions are stable in a part of the triple solution region, while the upper solution is always unstable (Barmak et al., 2016b). This is a region of operating conditions of particular interest, since in upward inclined flows this is the only region where stratified flow was experimentally observed (e.g., Barnea et al., 1980). Point **A** in Fig. 26 ($U_{1S} = -0.0003\,\text{m/s}$, $U_{2S} = -3.6\,\text{m/s}$), where both the lower holdup solution ($h = 0.05;\ \text{Re}_2 = 15.03;\ \text{Fr}_2 = 1.11 \cdot 10^{-3};\ \text{We}_2 = 5.47 \cdot 10^{-5}$) and the middle holdup solution ($h = 0.12;\ \text{Re}_2 = 18.9;\ \text{Fr}_2 = 2.24 \cdot 10^{-3};\ \text{We}_2 = 9.34 \cdot 10^{-5}$) are stable, is considered for studying the transient energy growth. This growth is very similar for both configurations and therefore only results for the lower holdup solution are shown in Fig. 27. It can be seen that even for small water flow rates a spanwise perturbation ($k_X = 0, k_Z = 4$) attains large energy growth ($G_{MAX}^{MAX} \approx 1700$). The transient energy growth of all studied perturbations lasts a short period of less than a second (Fig. 27 (b)), while the flow residence time in a long channel for the lower and middle holdup solutions is about 694 and 510 seconds, respectively. This is similar to the downward inclined flow case considered above, where the transient growth is rapid. It is dominated by the growth of the perturbation kinetic energy in the bulk of the thick air layer. The residence time based on the air velocity is $\tau_2 = 500H/|u_{2S}| = 2.8\text{s}$, and is of the order of the duration of the transient perturbation growth. It should be mentioned that the time step in the numerical calculation of the growth function and the optimal perturbation should be small enough (less than 0.01s) in order to correctly define the optimal perturbation characteristics.

While the energy growth is similar for both holdup solutions, the growth of the interface displacement amplitude is different for the lower holdup (Fig. 28(a)) and middle holdup (Fig. 28(b)) configurations. In the lower holdup solution, the interface displacement amplitude can reach a growth factor of more than three for an oblique perturbation with $k_X = 0.5, k_Z = 5$, while for the middle solution a growth factor of about nine was obtained for an oblique perturbation with $k_X = 0.5, k_Z = 4$. Nevertheless, due to strong non-modal growth of the kinetic energy, the interfacial (as well as gravitational) energy input to the total energy of the optimal perturbation is found to be negligibly small. Note that in the case of the middle holdup solution the optimal perturbation that attains $G_{MAX}^{MAX}$ (not shown) leads also to the growth of the interface displacement amplitude (up to a factor of three). This is in contrast to most of the other cases considered in this study, where the interface displacement amplitudes associated with global optimal perturbations tend to decay in time without initial growth. The provided analysis shows that the non-modal growth may lead to the flow pattern transition (most probably to stratified-wavy flow) for both the lower and middle holdup configurations of the triple-solution region.

## 5. CONCLUSIONS

The non-modal perturbation growth in two-phase stratified flow in horizontal and inclined channels under zero gravity and terrestrial conditions was studied. To elaborate effects of gravity in the total energy gain, the energy norm was extended to include kinetic, interfacial, and potential gravitational components of energy. Contrarily to the previous studies, we focused on the non-modal energy growth under linearly stable (i.e., subcritical) operational conditions.

We found that the largest energy gain is attained by a purely spanwise perturbation having a zero wavenumber in the streamwise direction. However, those perturbations do not exhibit growth of the interface displacement (an exception is the case of the middle holdup solution in upward inclined flow), while they exhibit significant growth in the streamwise velocity of the perturbation and in the kinetic energy. This process may be a precursor to transition process (on the route to turbulence) in one of the flow layers, similarly to single-phase shear flows (Schmid and Henningson, 2000). At the same time, we found that oblique optimal perturbations (with both non-zero streamwise and spanwise wavenumbers) yield the most noticeable deformation of the fluid-fluid interface.

Generally, in almost all of the cases considered, contribution of the interfacial energy is noticeable, so that its omission would lead to erroneous results. In particular, the growth function calculated via the kinetic energy norm only exhibits unphysical oscillations in time. The growth function calculated from the physically correct definition of the energy norm in most cases exhibits monotonically decay after initial growth (sometimes very short in time). Oscillation in the energy growth function can still occur and can be attributed to different types of instability modes (e.g., interfacial and shear modes) that can exchange energy during the growth process. On the other hand, the interface displacement amplitude always develops in time with oscillations that are characterized by the frequencies of the dominant modes.

In horizontal flows, the gravity affects the linear stability mainly through the modal mechanisms by expanding considerably the linearly stable region of flow rates as compared to that obtained under zero gravity. Consideration of the gravitational component of the energy in the non-modal analysis is essential for the correct predictions. However, its input has been found less significant than the interfacial and kinetic components, and more important for the flows with high density ratios. On the other hand, the energy gain in inclined flows was found to be substantial even for low liquid (heavy phase) flow rates corresponding to the linearly stable downward inclined air-water flows, and in the triple solution region of slightly upward inclined flows. As expected, in inclined flows the transient energy growth is affected significantly by the gravitational field. While in the test case of downward flow the growth of the interface displacement amplitude was found to be relatively small (less than in 2 times of the initial displacement), it is substantial (up to ten times) for both of the linearly stable configurations in the considered case of upward flow in the triple-solution region.

The conducted non-modal stability analysis has revealed that even for linearly stable (subcritical) conditions the transient energy growth can be considerable and large enough to trigger non-linear mechanisms of instability. While in some cases it can be associated with onset of shear mode of instability within one of the phases, in many cases there are optimal perturbations that were found to exhibit also noticeable- non-modal growth of the interface displacement amplitude. Such cases should be subject to a further study, where non-linear mechanisms are considered, as they may be associated with flow pattern transition and thereby with reduction of the region of operational conditions corresponding to stable smooth-stratified flow.

## APPENDIX: BASE FLOW

The base flow is assumed to be steady, laminar, and fully developed. Assuming that the velocity $U(y)$ is parallel to the channel walls and varies only with the cross-section coordinate $y$, the exact steady state solution can be found in the literature (e.g., Kushnir et al., 2014). The solution yields the steady (dimensionless) velocity profiles:

$$U_1 = c_1 + a_1 y + b_1 y^2 \text{ for } 0 \leq y_1 \leq 1, \quad U_2 = 1 + a_2 y + b_2 y^2 \text{ for } 0 \leq y_2 \leq 1, \tag{A1}$$

where

$$a_1 = \frac{n(a_2(n+1) + 2m + 2n - 2\tilde{Y})}{(n+1)m}, \quad a_2 = \frac{m - n^2 + n\tilde{Y}}{n^2 + n}, \quad b_1 = -\frac{n(m + n - \tilde{Y})}{(n+1)m}, \quad b_2 = -\frac{m + n + n\tilde{Y}}{n^2 + n},$$

$$\tilde{Y} = \frac{n(1-r)\text{Re}_2 \sin\beta}{2\text{Fr}_2} = \frac{Y(h-1)^2 (4h + m - 2mh + mh^2 - h^2)}{(Yh^4 - 3Yh^3 + 3Yh^2 - Yh - 1)}.$$

Given three dimensionless parameters $n$, $m$, $\tilde{Y}$, the base flow characteristics can be determined. However, since these parameters are based on the unknown lower (heavy) phase holdup, $h = h_1/H$, and the interfacial velocity $u_i$, it is convenient to use the other common parameters for two-phase flow, which are based on the specified operational conditions. The base flow solution is fully determined by the following three parameters: the Martinelli parameter $X^2 = (-dP/dx)_{1S}/(-dP/dx)_{2S} = m \cdot q$, the flow rate ratio $q = q_1/q_2$, and also the inclination parameter $Y = \rho_2 (r-1) g \sin\beta / (-dP/dx)_{2S}$. Here $q_j$ is the feed flow rate of phase $j$ and $(-dP/dx)_{jS} = 12\mu_j q_j / H^3$ is the superficial pressure drop corresponding to single phase flow in the channel, where $H = h_1 + h_2$.

The holdup can be found by solving the following algebraic equation $F(Y, q, m, h) = 0$ (e.g., Ullmann et al., 2003a):

$$Y - \frac{mq(1-h)^2 \left[(1+2h)m + (1-m)h(4-h)\right] - h^2 \left[(3-2h)m + (1-m)h^2\right]}{4h^3 (1-h)^3 \left[h + m(1-h)\right]} = 0. \tag{A2}$$

The interfacial velocity can be calculated as

$$\tilde{u}_i = \frac{u_i}{U_{2S}} = \frac{6h(1-h)(Yh - \tilde{P})}{m(1-h) + h}, \tag{A3}$$

where $U_{2S} = q_2 / H$ is the superficial velocity of the upper phase.

Note that Eq. (A2) yields a single solution for the holdup in horizontal flows, however it yields several stratified flow configurations (with different holdups) for fixed operational conditions in inclined flows (see details in Ullmann et al., 2003a, b). The feasibility of obtaining multiple holdups in inclined channels was validated experimentally (Ullmann et al., 2003a, b). Moreover, at least two distinct solutions were found to be stable in a range of superficial velocities in upward inclined concurrent and counter-current flow (Barmak et al., 2016b).

**REFERENCES**

Barmak, I., Gelfgat, A., Vitoshkin, H., Ullmann, A., Brauner, N., "Stability of stratified two-phase flows in horizontal channels," Phys. Fluids 28, 044101 (2016a).

Barmak, I., Gelfgat, A. Yu., Ullmann, A., Brauner, N., "Stability of stratified two-phase flows in inclined channels," Phys. Fluids 28, 084101 (2016b).

Barmak, I., Gelfgat, A. Yu., Ullmann, A., Brauner, N., "On the Squire's transformation for stratified two-phase flows in inclined channels," Int. J. Multiphase Flow 88, 142–151 (2017).

Barnea, D., Shoham, O., Taitel, Y., Dukler, A. E., "Flow pattern transition for gas-liquid flow in horizontal and inclined pipes," Int. J. Multiphase Flow 6, 217-225 (1980).

Barnea, D., Shoham, O., Taitel, Y., "Flow pattern transition for downward inclined two phase flow; horizontal to vertical," Chem. Eng. Sci. 37, 735-740 (1982).

Brandt, L., "The lift-up effect: The linear mechanism behind transition and turbulence in shear flows," Eur. J. Mech. B/Fluids 47, 80-96 (2014).

Charru, F., Fabre, J., "Long waves at the interface between two viscous fluids," Phys. Fluids 6, 1223-1235 (1994).

Farrell, B. F., "Optimal excitation of perturbations in viscous shear flow," Phys. Fluids 31, 2093-2102 (1988).

Gustavsson, L. H., "Energy growth of three-dimensional disturbances in plane Poiseuille flow," J. Fluid Mech. 224, 241-260 (1991).

Hooper, A. P., Boyd, W. S. G., "Shear-flow instability at the interface between two viscous fluids," J. Fluid Mech. 128, 507-528 (1983).

Jose, S., Kuzhimparampil, V., Pier, B., Govindarajan, R., "Algebraic disturbances and their consequences in rotating channel flow transition," Phys. Rev. Fluids 2, 083901 (2017).

Kaffel, A., Riaz, A., "Eigenspectra and mode coalescence of temporal instability in two-phase channel flow." Phys. Fluids 27, 042101 (2015).

Kushnir, R., Segal, V., Ullmann, A., Brauner, N., "Inclined two-layered stratified channel flows: Long wave stability analysis of multiple solution regions," Int. J. Multiphase Flow 62, 17-29 (2014).

Lucas, J.-M., Vermeersch, O., Arnal, D., "Transient growth of Görtler vortices in two-dimensional compressible boundary layers. Application to surface waviness," Eur. J. Mech. B/Fluids 50, 132-146 (2015).

Malik, S., Hooper, A. P., "Three-dimensional disturbances in channel flows," Phys. Fluids 19, 052102 (2007).

Ó Náraigh, L., Valluri, P., Scott, D. M., Bethune, I., Spelt, P. D. M., "Linear instability, nonlinear instability and ligament dynamics in three-dimensional laminar two-layer liquid-liquid flows," J. Fluid Mech. 750, 464-506 (2014).

Orszag, S. A., "Accurate solution of the Orr-Sommerfeld stability equation," J. Fluid Mech. 50, 689-703 (1971).

Reddy, S. C., Henningson, D. S., "Energy growth in viscous channel flows," J. Fluid Mech. 252, 209-238 (1993).

Renardy, Y., "The thin layer effect and interfacial stability in a two-layer Couette flow with similar liquids," Phys. Fluids 30, 1627–1637 (1987).

Schmid, P. J., Henningson, D. S., *Stability and Transition in Shear Flows* (Springer, New York, 2000).

South, M. J., Hooper, A. P., "Linear growth in two-fluid plane Poiseuille flow," J. Fluid Mech. 381, 121–139 (1999).

Tilley, B.S., Davis, S.H., Bankoff, S.G., "Linear stability of two-layer fluid flow in an inclined channel," Phys. Fluids 6, 3906–3922 (1994).

Ullmann, A., Zamir, M., Ludmer, Z., Brauner, N., "Stratified laminar countercurrent flow of two liquid phases in inclined tube," Int. J. Multiphase Flow 29, 1583–1604 (2003a).

Ullmann, A., Zamir, M., Gat, S., Brauner, N., "Multi-holdups in co-current stratified flow in inclined tubes," Int. J. Multiphase Flow 29, 1565–1581 (2003b).

van Noorden, T. L., Boomkamp, P. A. M., Knaap, M. C., Verheggen, T. M. M., "Transient growth in parallel two-phase flow: Analogies and differences with single-phase flow," Phys. Fluids 10, 2099–3001 (1998).

Vitoshkin, H., Heifetz, E., Gelfgat, A. Yu., Harnik, N., "On the role of vortex stretching in energy optimal growth of three-dimensional perturbations on plane parallel shear flows", J. Fluid Mech. 707, 369-380 (2012).

Yecko, P., Zaleski, S., "Transient growth in two-phase mixing layers," J. Fluid Mech. 528, 43–52 (2005).

Yecko, P., "Disturbance growth in two-fluid channel flow: the role of capillarity," Int. J. Multiphase Flow 34, 272-282 (2008).

Yiantsios, S. G., Higgins, B. G., "Linear stability of plane Poiseuille flow of two superposed fluids," Phys. Fluids 31, 3225–3238 (1988).

Yih, C. S., "Instability due to viscosity stratification," J. Fluid Mech. 27, 337–352 (1967).

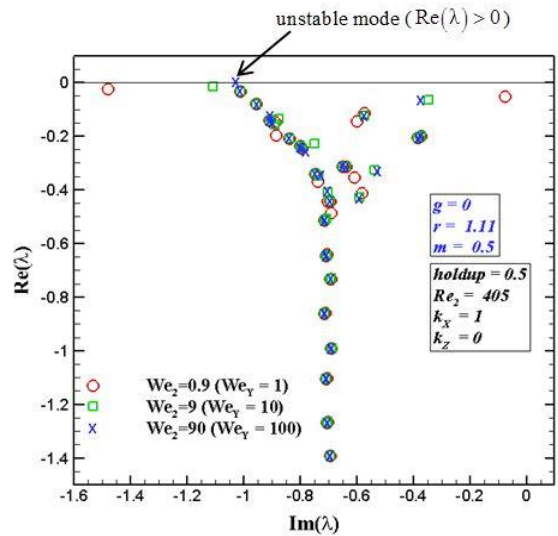

FIG. 2. Spectra of zero-gravity systems with different values of surface tension (Weber number) for a 2D perturbation $(k_x = 1, k_z = 0)$: $r = 1.11$, $m = 0.5$, $Re_2 = 405$, $h = 0.5$ (YC). The results are identical to those presented in Fig. 2a of Yecko (2008).

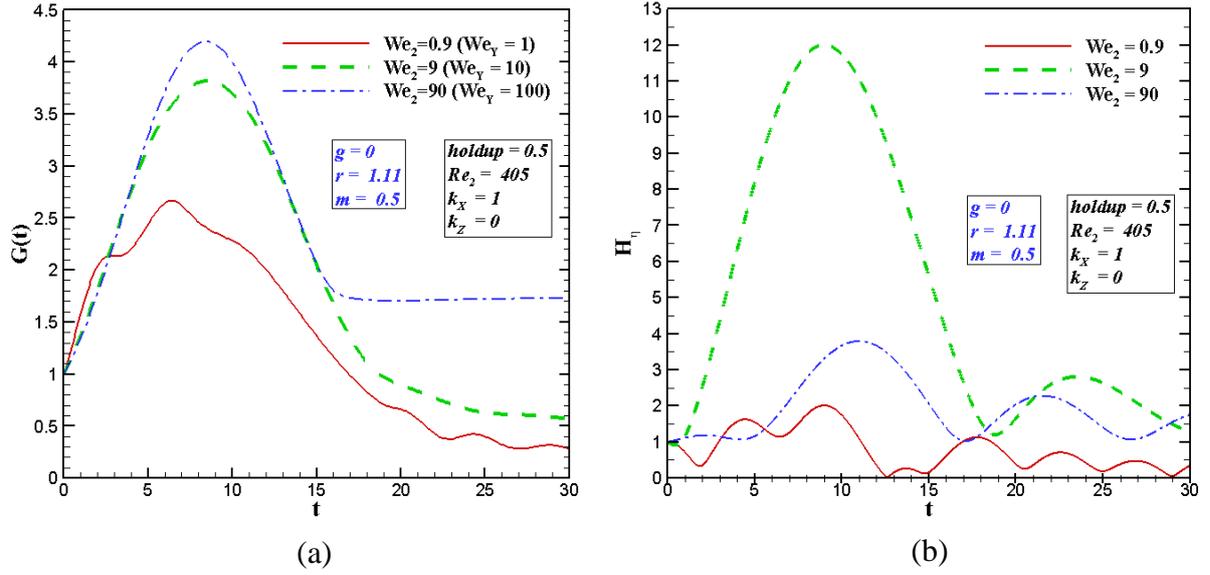

FIG. 3. (a) Energy growth function vs. dimensionless time, $G(t)$, for $k_X = 1, k_Z = 0$ and (b) time evolution of the interface displacement amplitude corresponding to the optimal perturbation (associated with a maximum of $G(t)$, $G_{MAX}$) for YC (with different Weber numbers).

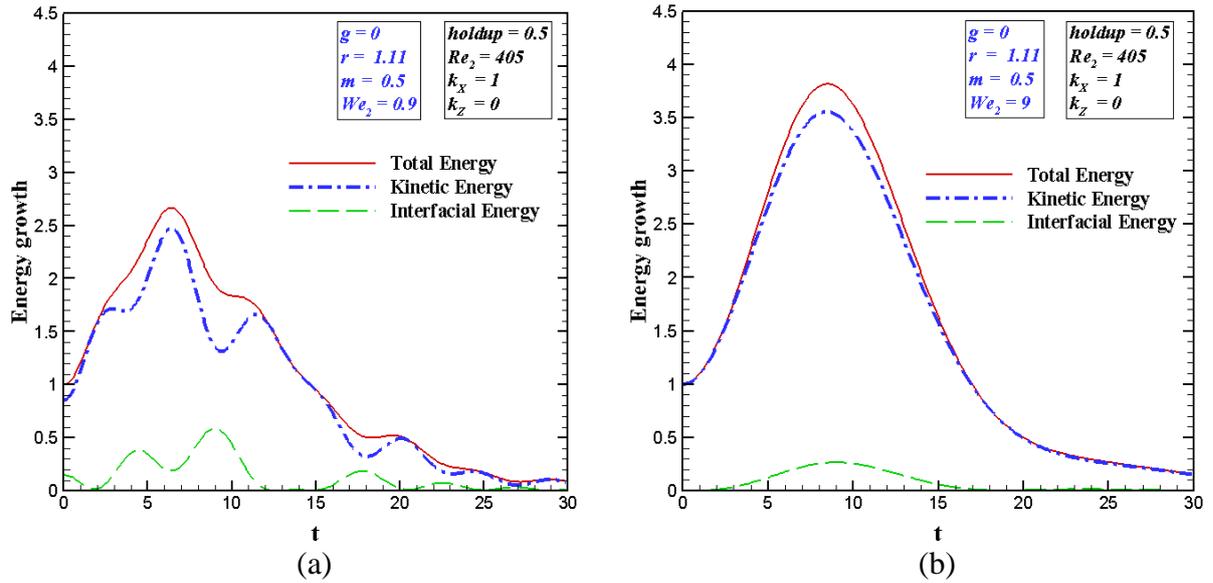

FIG. 4. Time evolution of energy of the optimal perturbation for YC for (a) $We_2 = 0.9$ and (b) $We_2 = 9$. All the components of energy are normalized by a value of the total energy at $t = 0$.

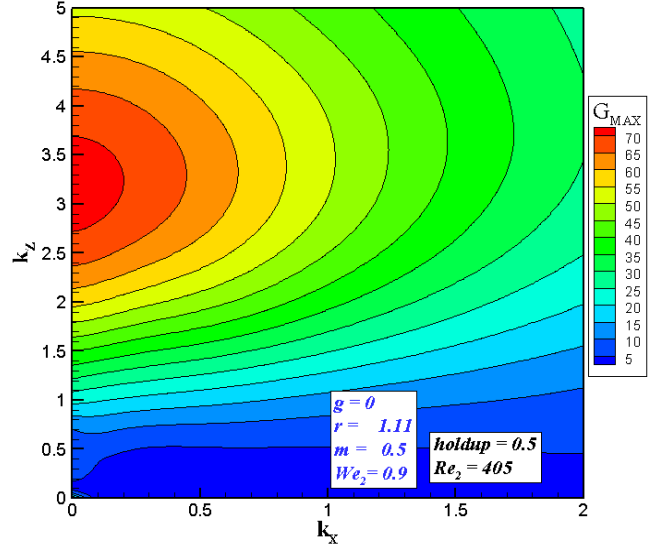

FIG. 5. Contours of maximal energy growth, $G_{MAX}(k_X, k_Z)$, for YC, $We_2 = 0.9$.

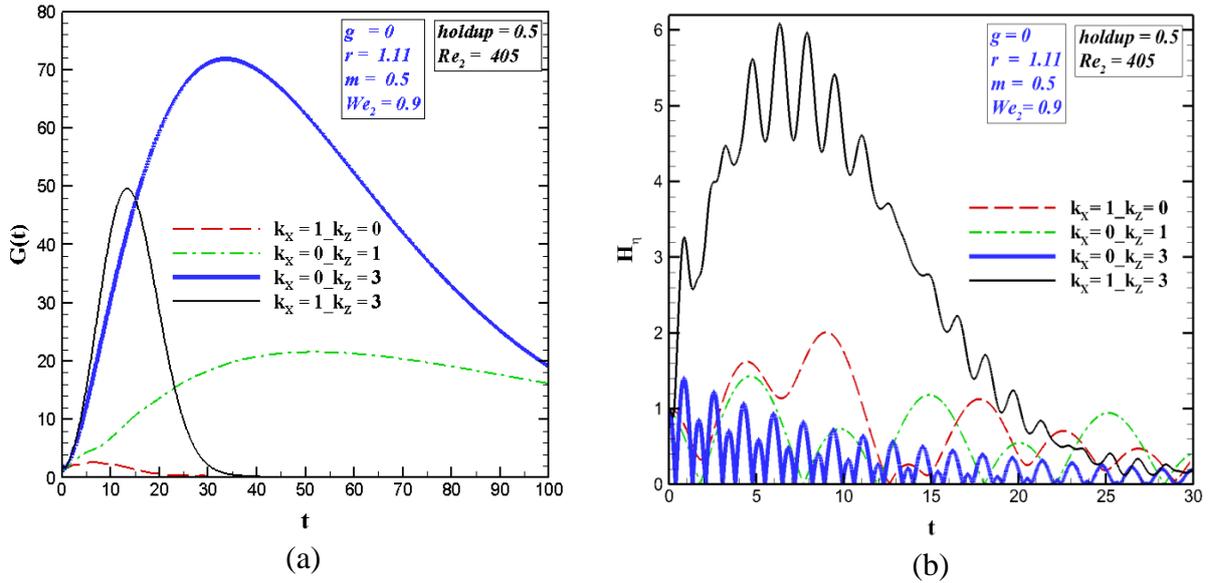

FIG. 6. (a) Energy growth function, $G(t)$; (b) time evolution of the interface displacement amplitude of the corresponding optimal perturbations for YC, $We_2 = 0.9$.

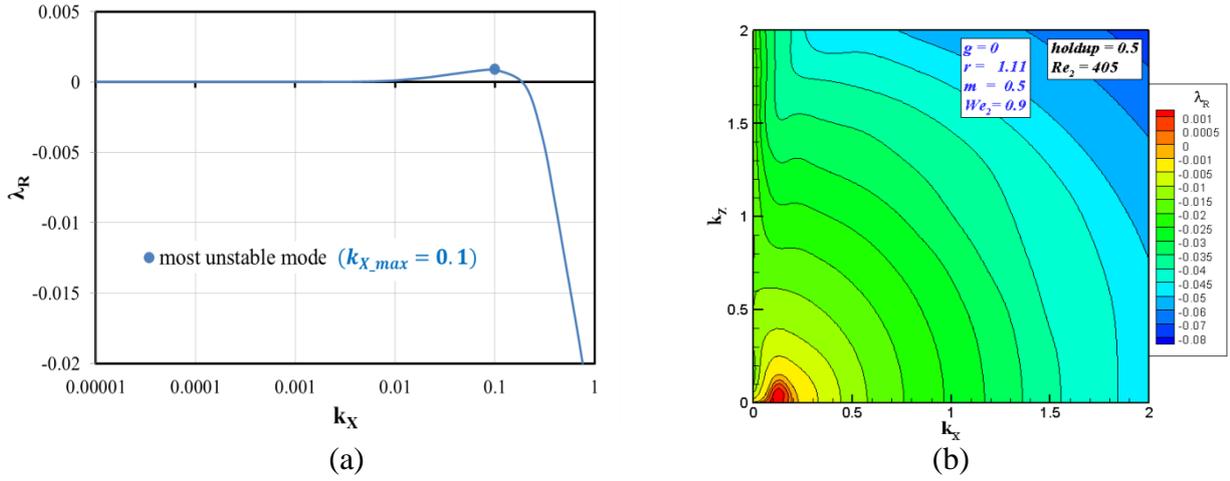

FIG. 7. Modal stability analysis (exponential growth/decay of perturbations) for YC, $We_2 = 0.9$: (a) growth rate $(\lambda_R)$ for 2D perturbations; (b) contours of growth rate, $\lambda_R(k_X, k_Z)$.

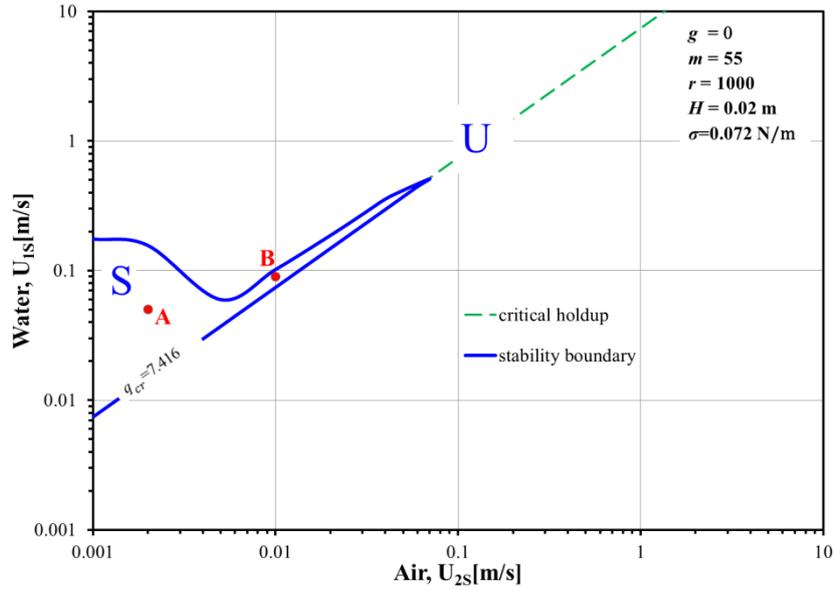

FIG. 8. Stability map for air-water flow under zero-gravity conditions. **S** and **U** denote the modally stable and unstable regions, respectively. Points **A** $(h = 0.95, U_{2S} = 0.002\,\text{m/s}, U_{1S} = 0.05\,\text{m/s}; \text{Re}_2 = 49.5;$ $\text{Fr}_2 = 8.26 \cdot 10^{-2}; We_2 = 1.12 \cdot 10^{-3})$ and **B** $(h = 0.89, U_{2S} = 0.01\,\text{m/s}, U_{1S} = 0.09\,\text{m/s}; \text{Re}_2 = 17.4; \text{Fr}_2 = 1.08;$ $We_2 = 6.58 \cdot 10^{-4})$ are selected for the non-modal analysis.

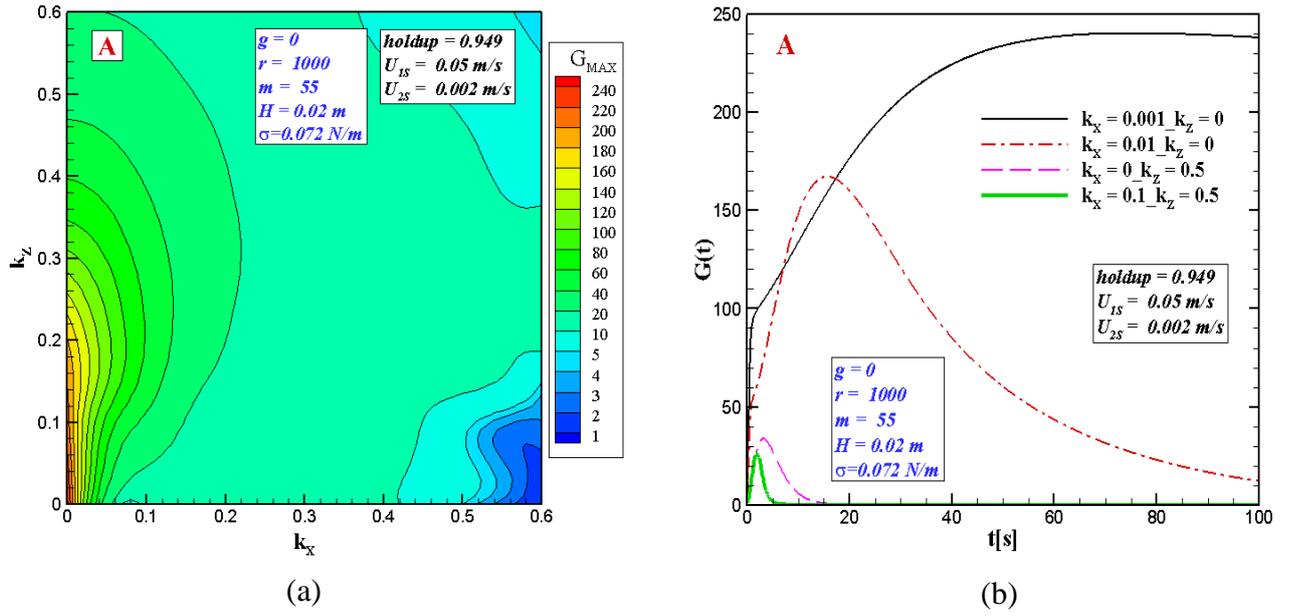

FIG. 9. (a) Contours of maximal energy growth, $G_{MAX}(k_X, k_Z)$; (b) energy growth function $G(t)$ for point **A** (Fig. 8). A characteristic "residence time" in a long channel of $500H$ at point **A** is $\tau = 500H/u_i = 135s$.

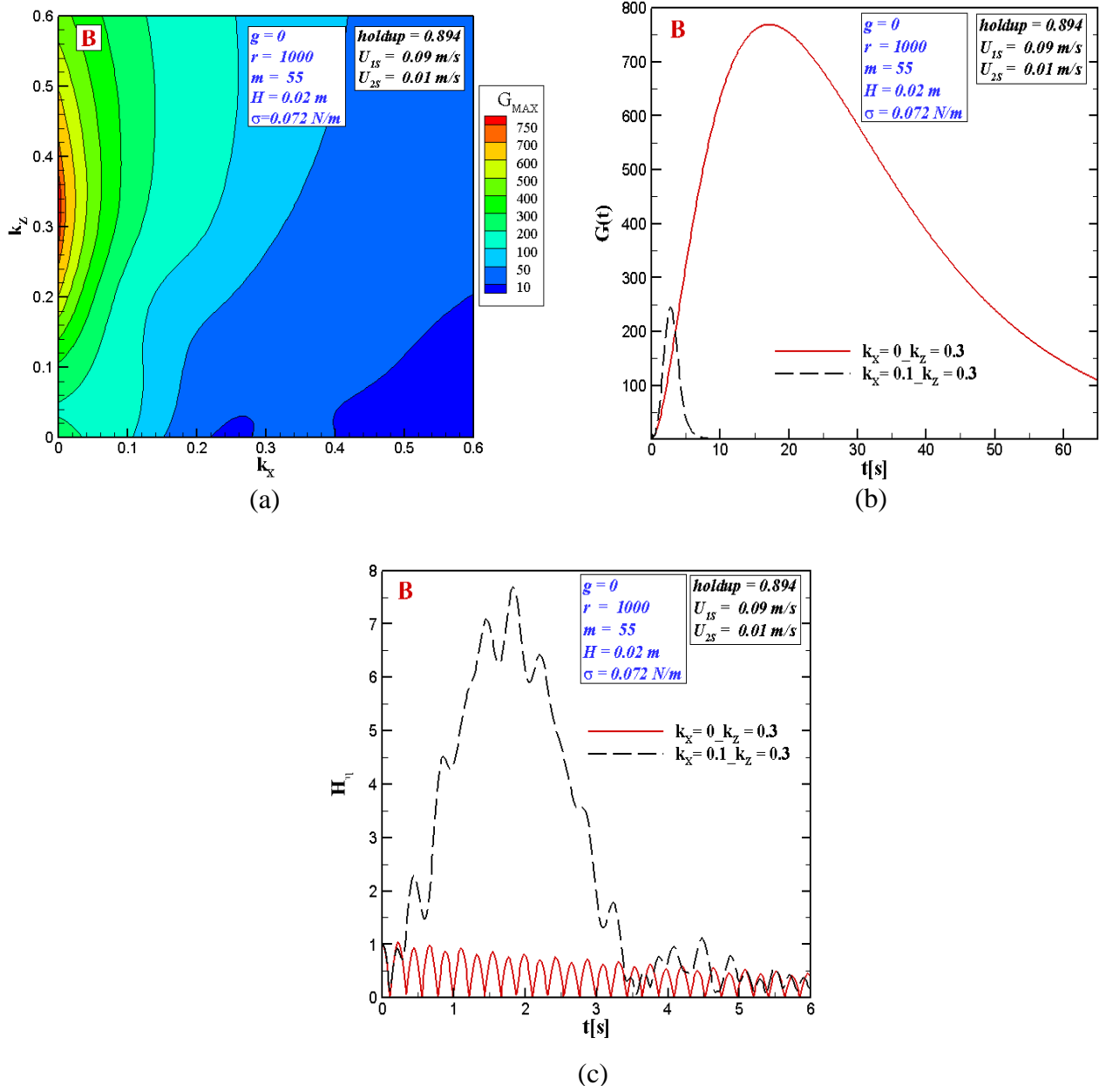

FIG. 10. (a) Contours of maximal energy growth, $G_{MAX}(k_X, k_Z)$; (b) energy growth function $G(t)$ for point **B** in Fig. 8; (c) time evolution of the interface displacement amplitude of the corresponding optimal perturbations. A characteristic "residence time" at point **B** is $\tau = 500H/u_i = 67$s.

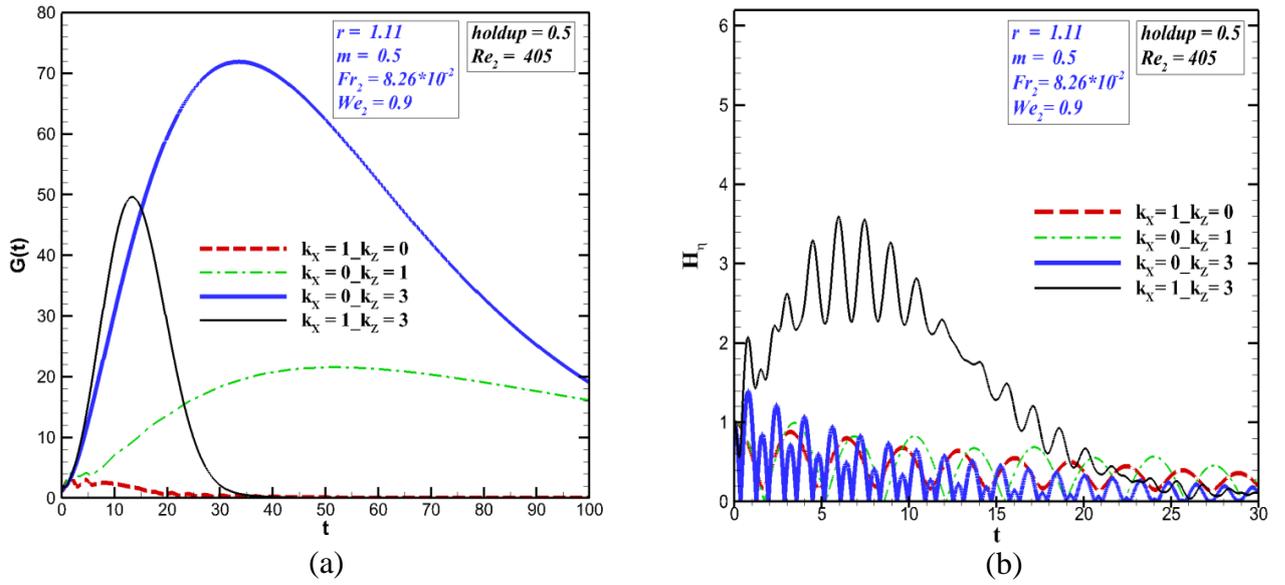

FIG. 11. (a) Energy growth function, $G(t)$; (b) time evolution of the interface displacement amplitude of the corresponding optimal perturbations for the normal gravity system with the same parameters as in YC.

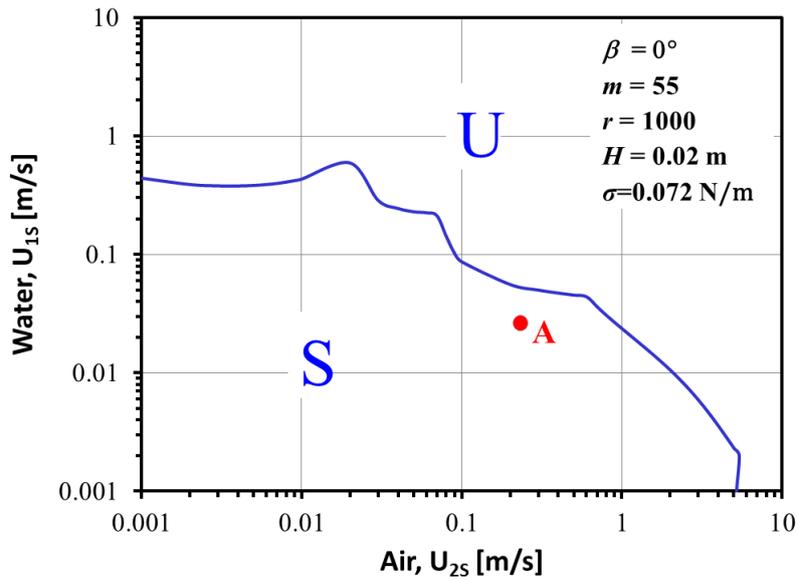

FIG. 12. Stability map for air-water flow in a 2 cm horizontal channel. **S** and **U** denote the modally stable and unstable regions, respectively. Point **A** $(h = 0.5;\ U_{2S} = 0.2325\,\text{m/s};\ U_{1S} = 0.026318\,\text{m/s}$ $\text{Re}_2 = 49.5;\ \text{Fr}_2 = 8.26\cdot 10^{-2};\ \text{We}_2 = 1.12\cdot 10^{-3})$ is selected for the non-modal analysis.

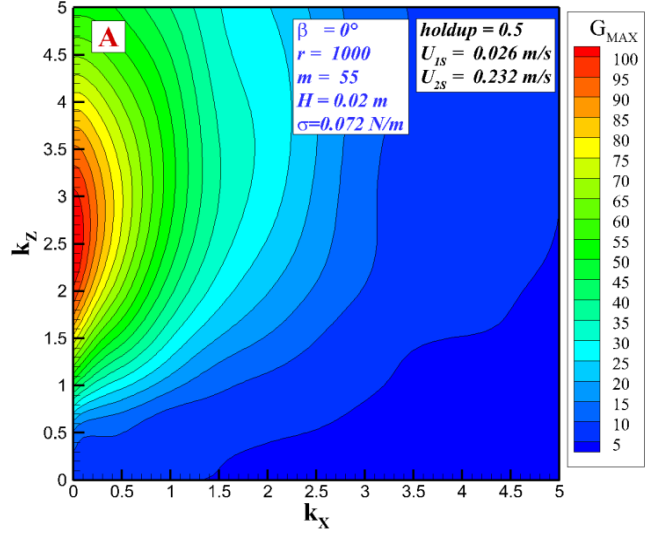

FIG. 13. Contours of maximal transient energy growth, $G_{MAX}(k_x, k_z)$, for point **A** (Fig. 12).

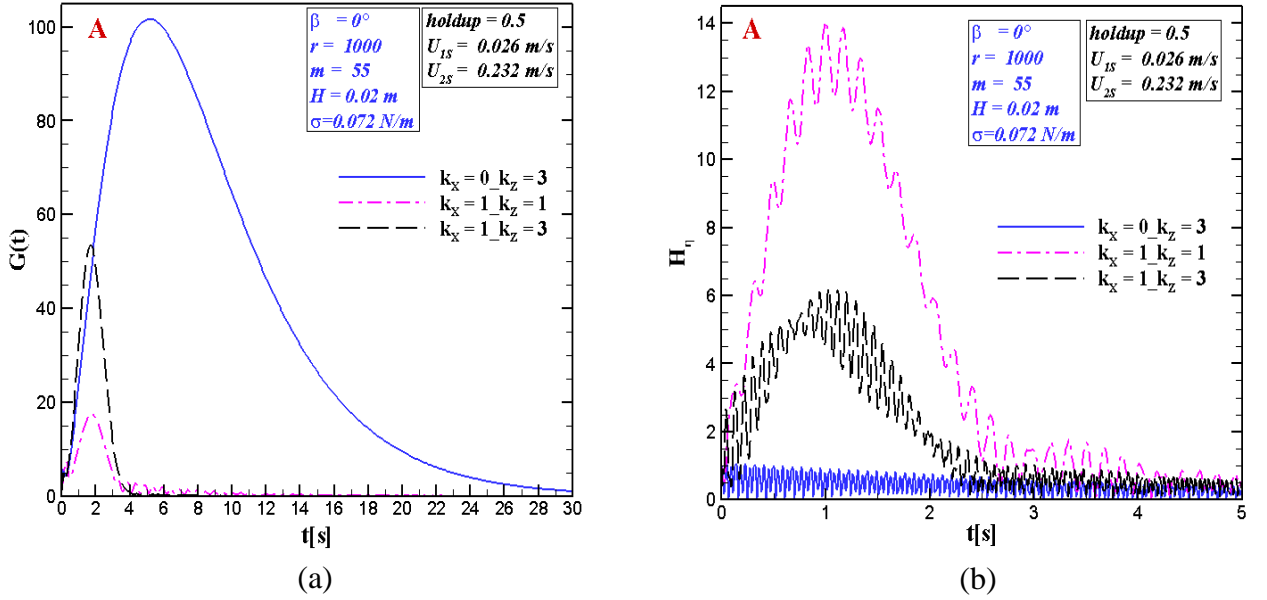

(a)  (b)

FIG. 14. (a) Energy growth function, $G(t)$; (b) time evolution of the interface displacement amplitude of the corresponding optimal perturbations for point **A** (Fig. 12). A characteristic "residence time" at point **A** is $\tau = 500H/u_i = 111\text{s}$.

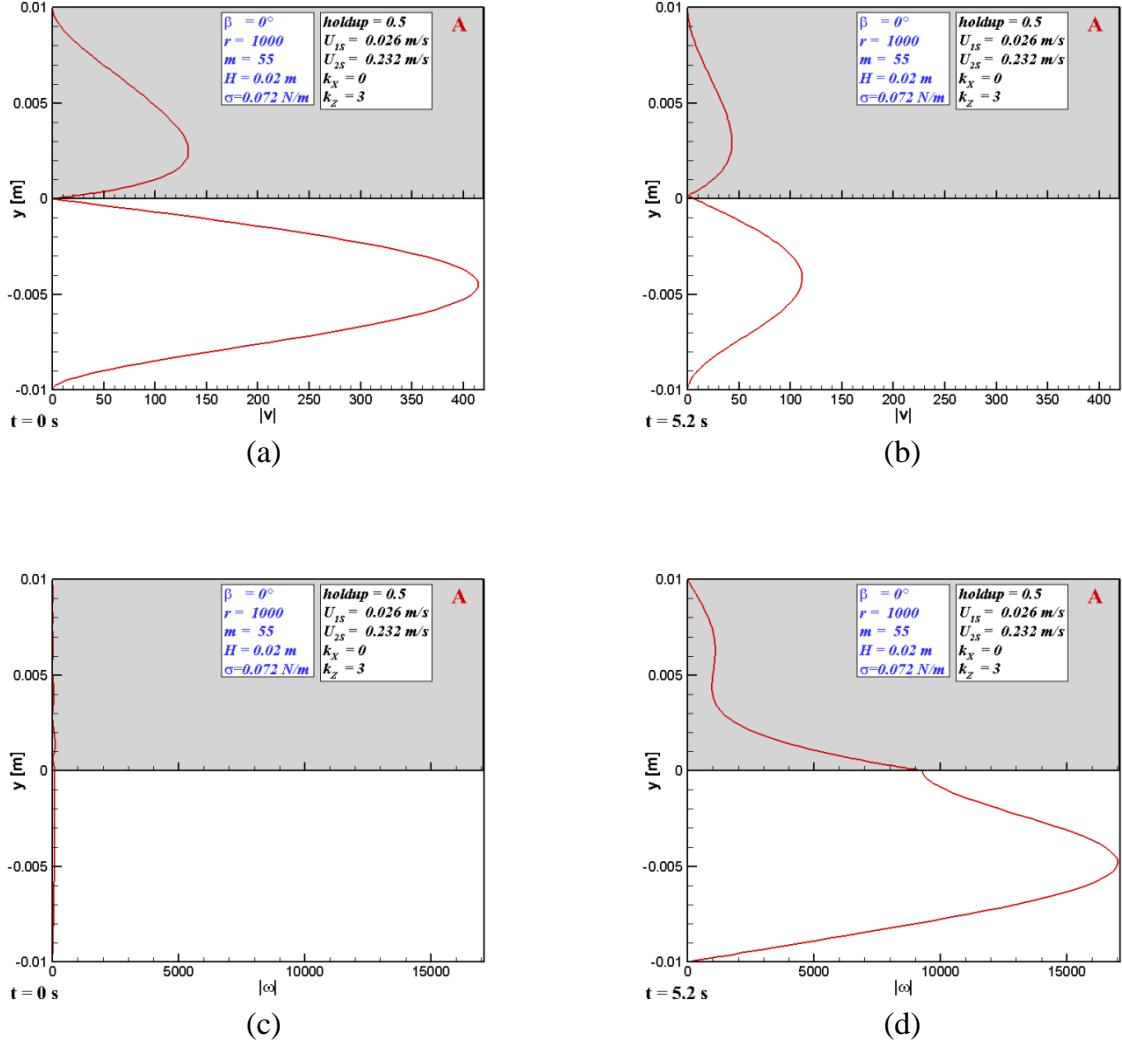

FIG. 15. Cross-section distribution of the optimal perturbation ($k_X = 0, k_Z = 3$) for point **A** (Fig. 12): absolute value of the transverse velocity amplitude $|v|$ (a) $t = 0$; (b) $t = 5.2\,\text{s}$ (at which $G(t) = G_{MAX}^{MAX}$); absolute value of the y-vorticity amplitude (c) $t = 0$; (d) $t = 5.2\,\text{s}$ (at which $G(t) = G_{MAX}^{MAX}$). Values of $|v|, |\omega|$ are normalized by the value of $|v|$ at the interface at $t = 0$ ($|v(y=0, t=0)| = 1$; $|\omega(y=0, t=0)| \approx 100$).

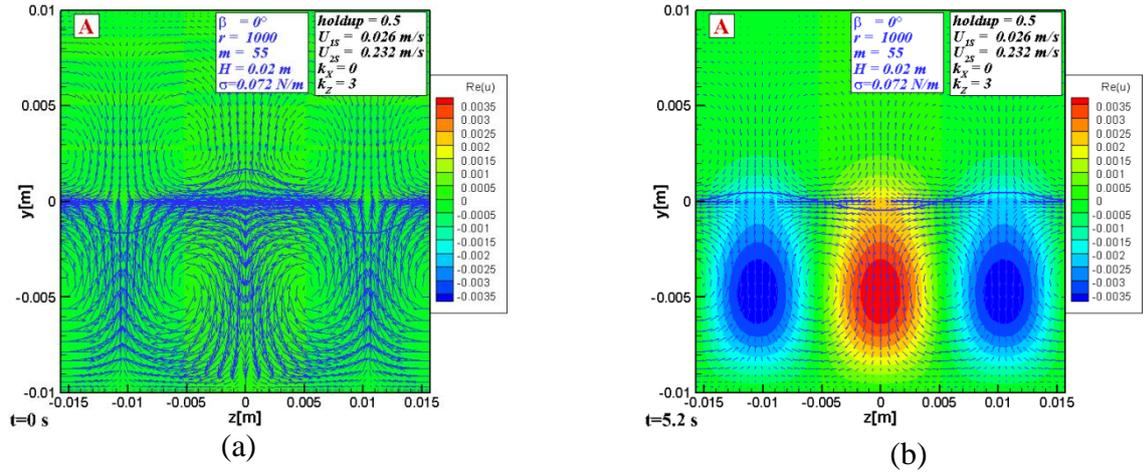

FIG. 16. The $(v, w)$ vectors and the $u$ - contours of the optimal perturbation for point **A** (Fig. 12), $k_X = 0, k_Z = 3$: (a) $t = 0$ and (b) $t = 5.2[s]$ (at which $G(t) = G_{MAX}^{MAX}$); blue solid line depicts the deformed interface.

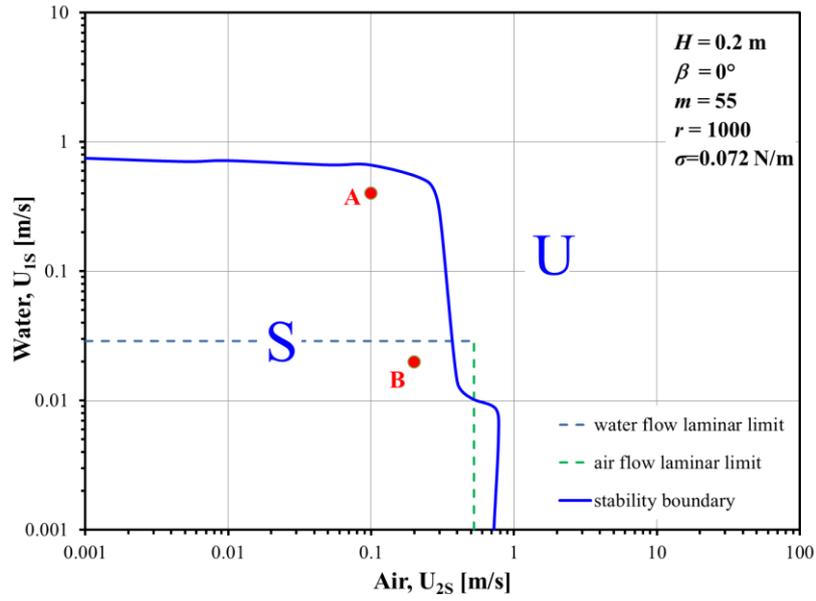

FIG. 17. Stability map for air-water flow in a 20 cm horizontal channel. **S** and **U** denote the modally stable and unstable regions, respectively. Points **A** $(h = 0.83; U_{2S} = 0.1\,\text{m/s}; U_{1S} = 0.4\,\text{m/s}; \text{Re}_2 = 1332.1;$ $\text{Fr}_2 = 1.66; \text{We}_2 = 0.25)$ and **B** $(h = 0.49, U_{2S} = 0.2\,\text{m/s}, U_{1S} = 0.02\,\text{m/s}; \text{Re}_2 = 397.5; \text{Fr}_2 = 4.93 \cdot 10^{-3};$ $\text{We}_2 = 7.07 \cdot 10^{-3})$ are selected for the non-modal analysis.

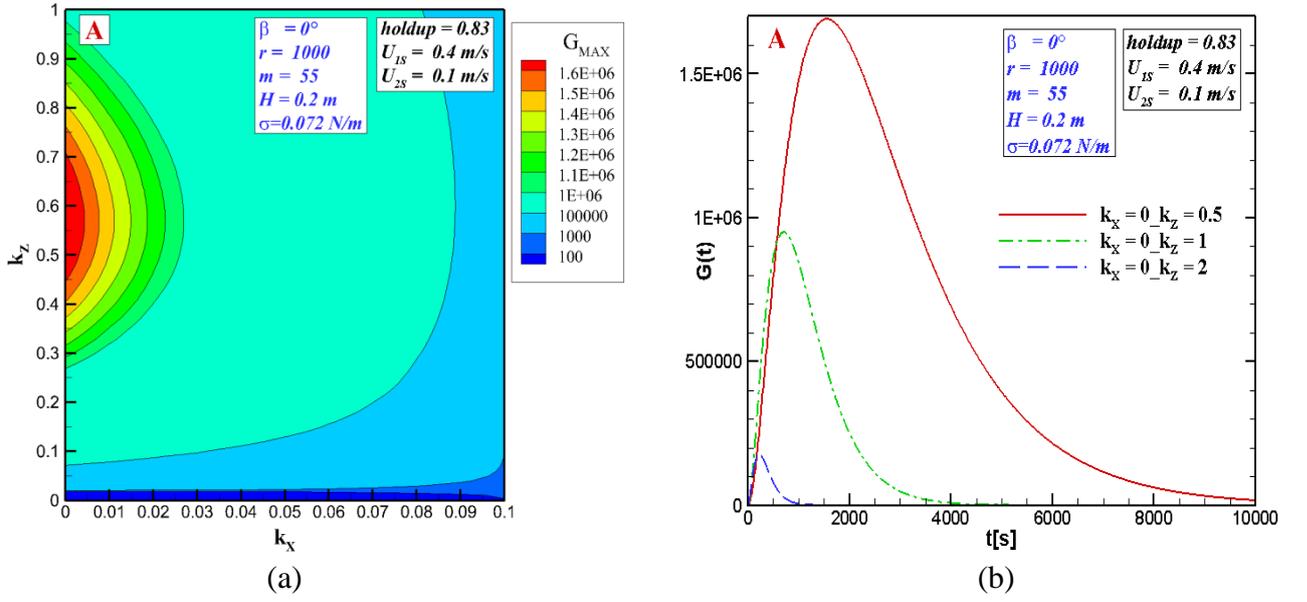

FIG. 18. (a) Contours of maximal transient energy growth, $G_{MAX}(k_X, k_Z)$; (b) energy growth function, $G(t)$, for point **A** (Fig. 17). A characteristic "residence time" at point **A** is $\tau = 500H/u_i = 136s$.

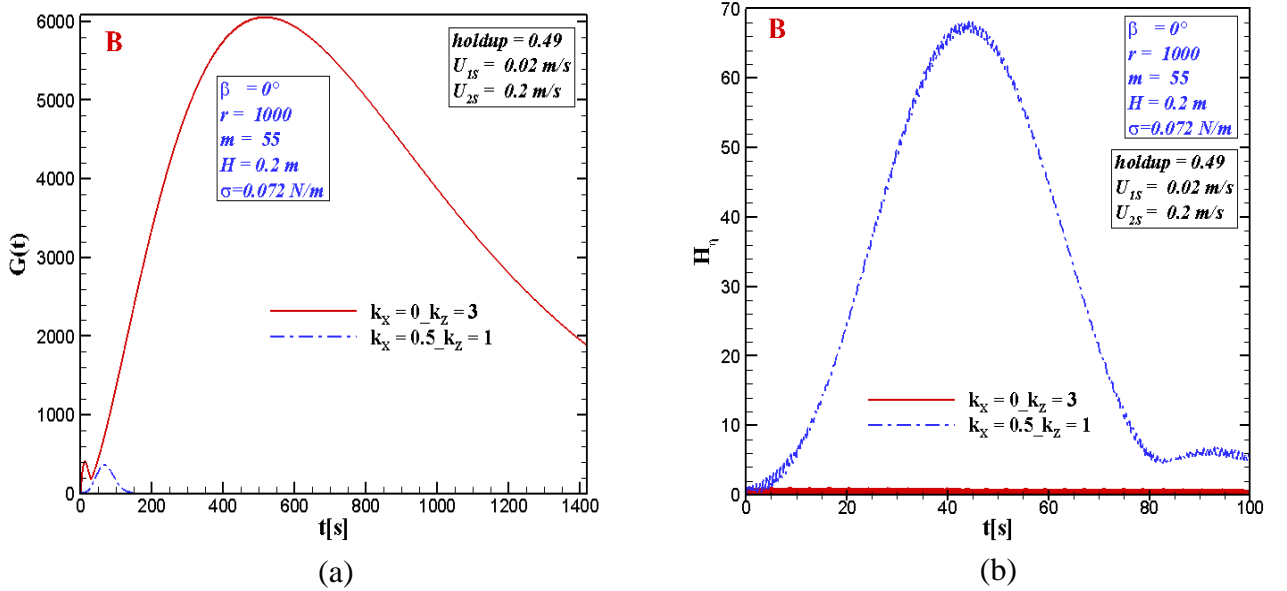

FIG. 19. (a) Energy growth function, $G(t)$; (b) Time evolution of the interface displacement amplitude of the corresponding optimal perturbations for point **B** (Fig. 17). $H_\eta$ keeps decaying at longer times. A characteristic "residence time" at point **B** is $\tau = 500H/u_i = 1419s$.

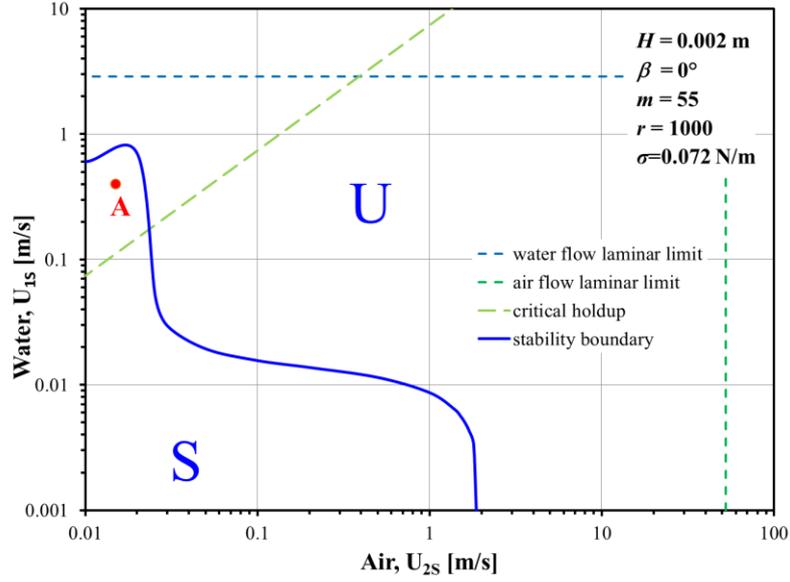

FIG. 20. Stability map for air-water flow in a 2 mm horizontal channel. **S** and **U** denote the modally stable and unstable regions, respectively. Point **A** $\left(h=0.95; U_{2S}=0.015\,\text{m/s}; U_{1S}=0.4\,\text{m/s}; \text{Re}_2=3.11; \text{Fr}_2=365.04; \text{We}_2=4.61\cdot 10^{-4}\right)$ is selected for the non-modal analysis.

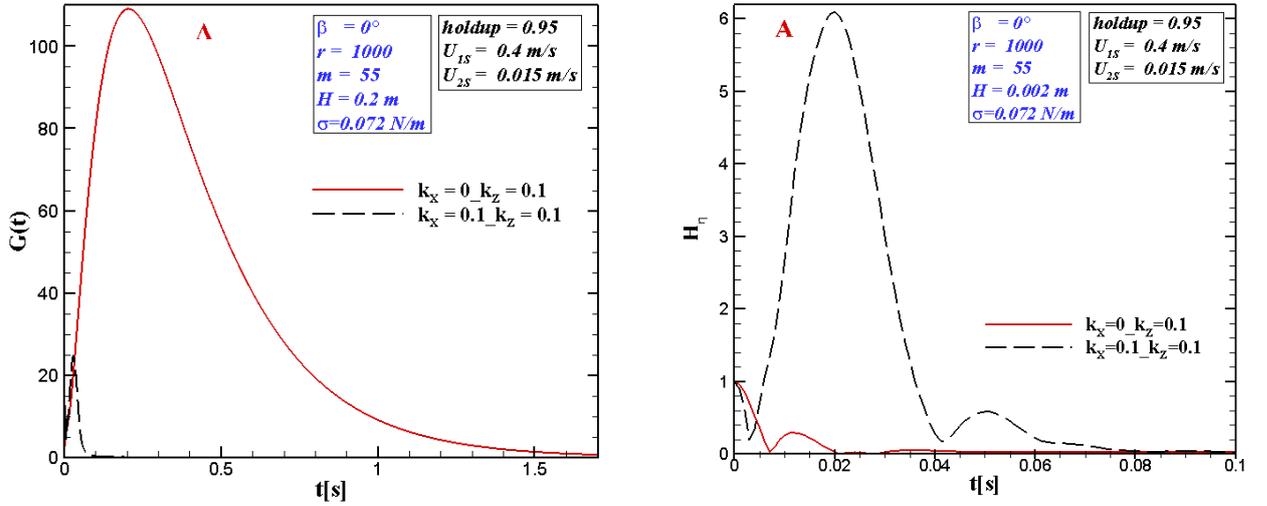

FIG. 21. (a) Energy growth function, $G(t)$; (b) time evolution of the interface displacement amplitude of the corresponding optimal perturbations for point **A** (Fig. 20). $H_\eta$ keeps decaying at longer times. A characteristic "residence time" at point **A** is $\tau = 500 H/u_i = 1.7\,\text{s}$.

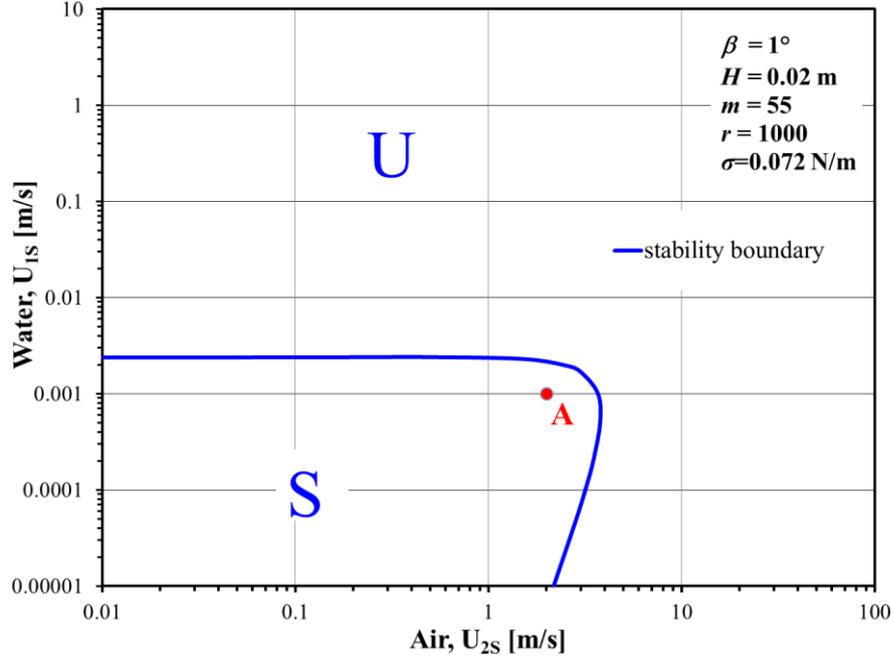

FIG. 22. Stability map for downward inclined air-water flow. **S** and **U** denote the modally stable and unstable regions, respectively. Point **A** $\left(h=0.03; U_{2S}=2\,\text{m/s}; U_{1S}=0.001\,\text{m/s}; \text{Re}_2=49.5;\right.$ $\left.\text{Fr}_2=1.14\cdot10^{-2}; \text{We}_2=5.83\cdot10^{-4}\right)$ is selected for the non-modal analysis.

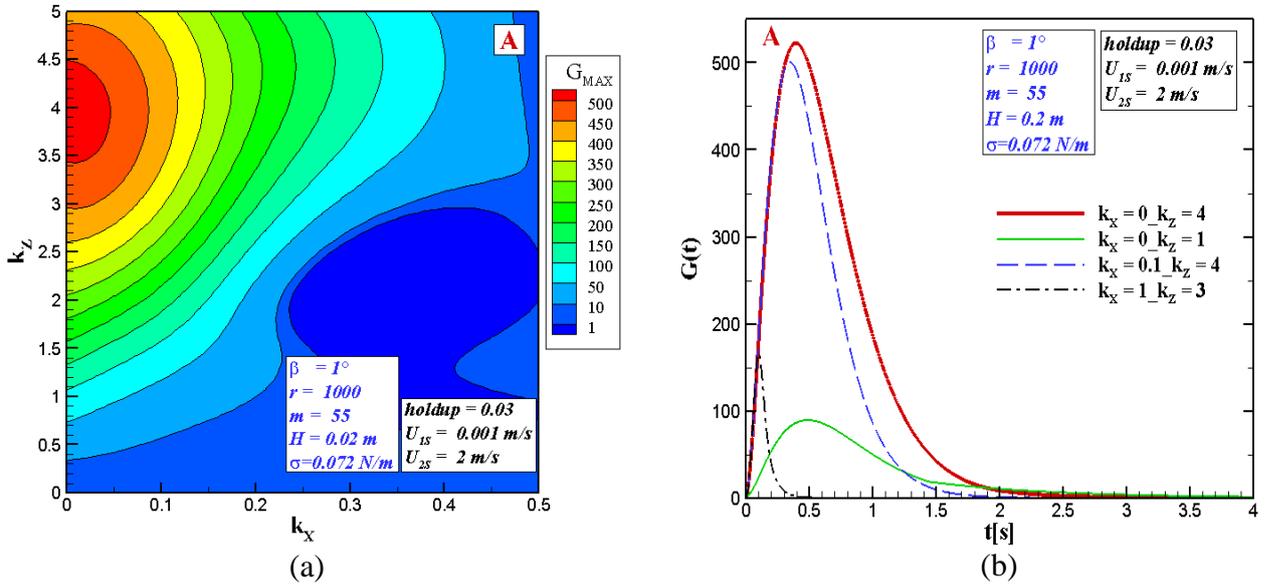

(a)          (b)

FIG. 23. (a) Contours of maximal transient energy growth, $G_{MAX}(k_X, k_Z)$; (b) energy growth function, $G(t)$, for point **A** (Fig. 22). A characteristic "residence time" at point **A**, $\tau = 500H/u_i = 214\text{s}$, the air phase residence time - $\tau_2 = 500H/u_{2S} = 5\text{s}$.

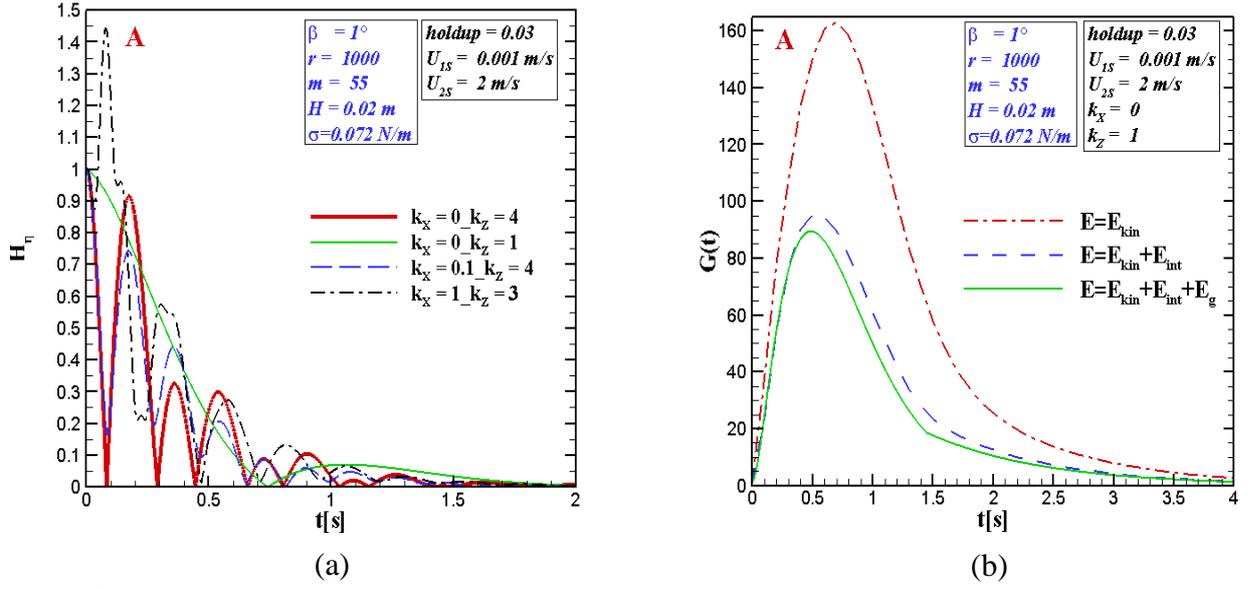

FIG. 24. (a) Time evolution of the interface displacement amplitude of the corresponding optimal perturbations for point **A** (Fig. 22); (b) Energy growth function, $G(t)$, obtained from different energy norm definitions (that take into account only part of the components) for point **A** and $k_X = 0, k_Z = 1$. $G(t)$ is scaled by a value of the corresponding energy norm at $t = 0$. A characteristic "residence time" at point **A** is $\tau = 500H/u_i = 214\text{s}$, the air phase residence time - $\tau_2 = 500H/u_{2S} = 5\text{s}$.

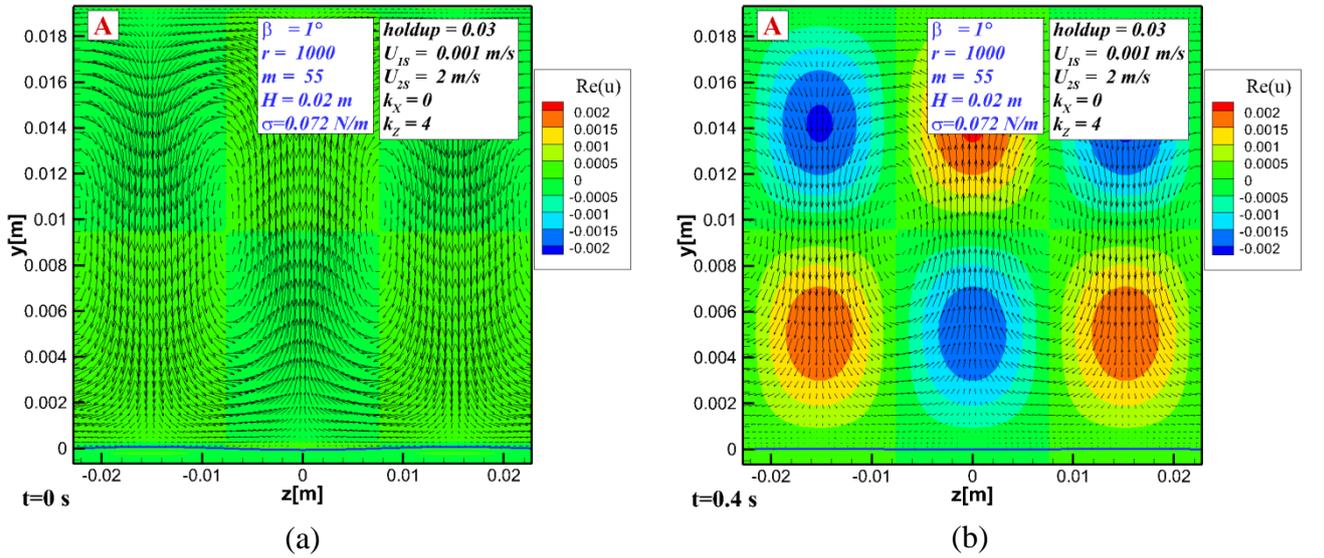

FIG. 25. The $(v, w)$ vectors and the $u$ - contours of the optimal perturbation for point **A** (Fig. 22), $k_X = 0, k_Z = 4$: (a) $t = 0$ and (b) $t = 0.4[\text{s}]$ (at which $G(t) = G_{MAX}^{MAX}$); blue solid line depicts the deformed interface.

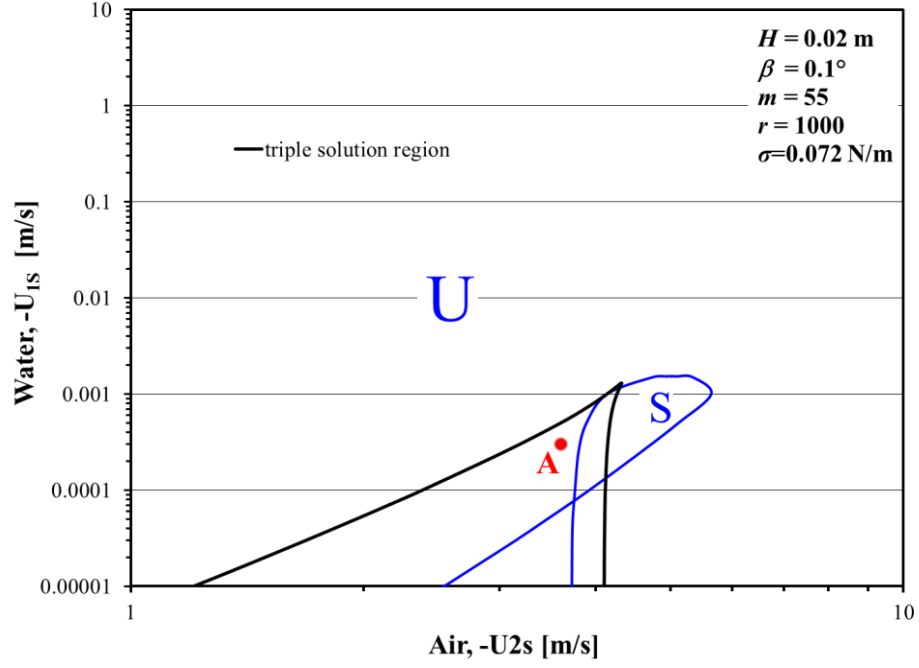

FIG. 26. Stability map for downward inclined air-water flow in the triple solution region. **S** and **U** denote the modally stable and unstable regions, respectively. Point **A** $\left(U_{2S}=-3.6\,\text{m/s}; U_{1S}=-0.0003\,\text{m/s}\right)$ is selected for the non-modal analysis.

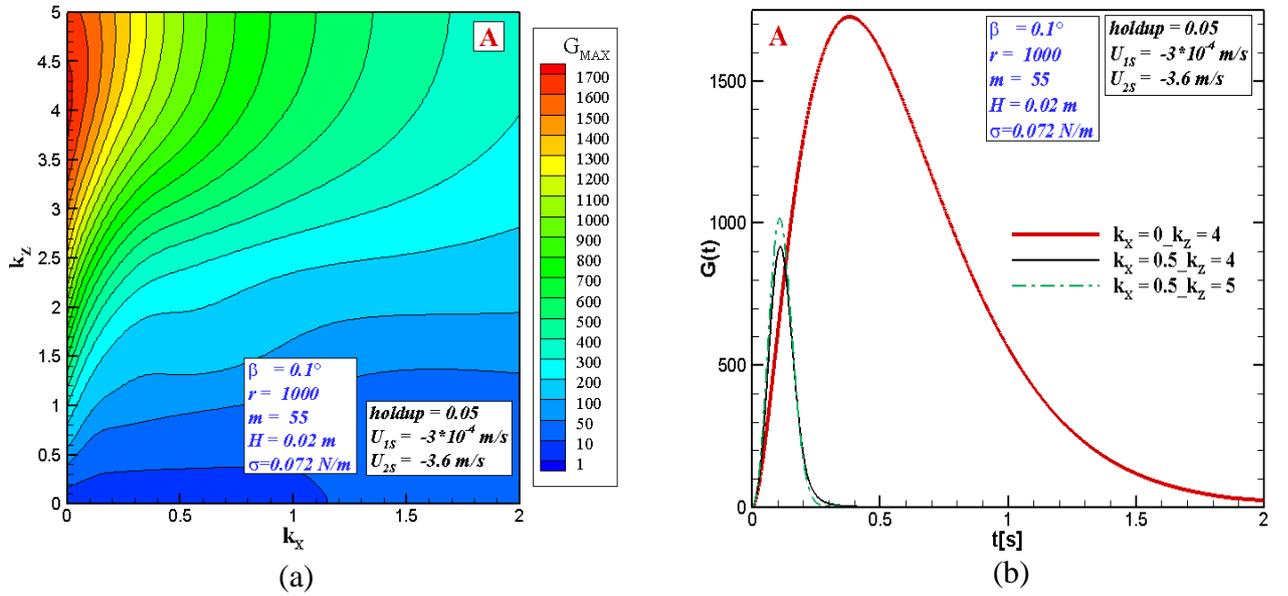

FIG. 27. (a) Contours of maximal transient energy growth, $G_{MAX}(k_X, k_Z)$ (b) Energy growth function, $G(t)$, for the lower holdup solution $\left(h=0.05; \text{Re}_2=15.03; \text{Fr}_2=1.11\cdot10^{-3}; \text{We}_2=5.47\cdot10^{-5}\right)$ at point **A** (Fig. 26). A characteristic "residence time" for the lower holdup solution at point **A** is, $\tau=500H/|u_i|=694\text{s}$, the air phase residence time - $\tau_2=500H/|u_{2S}|=2.8\text{s}$.

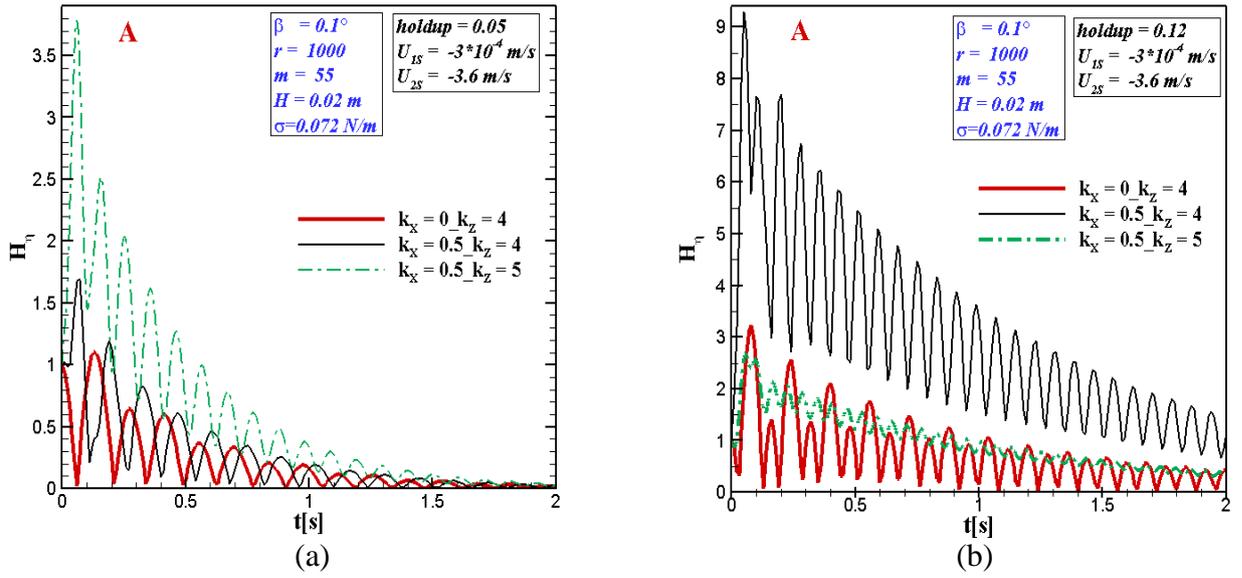

FIG. 28. Time evolution of the interface displacement amplitude of the several optimal perturbations for point **A** (Fig. 26): (a) lower holdup solution; (b) middle holdup solution $\left(h = 0.12; \text{Re}_2 = 18.9; \text{Fr}_2 = 2.24 \cdot 10^{-3}; \text{We}_2 = 9.34 \cdot 10^{-5}\right)$. A characteristic "residence time" for the middle holdup solution at point **A** is $\tau = 500H/|u_i| = 510\text{s}$, the air phase residence time - $\tau_2 = 500H/|u_{2S}| = 2.8\text{s}$.